\input harvmac 
\input epsf.tex
\def\IN{\relax{\rm I\kern-.18em N}} 
\def\IR{
\relax{\rm I\kern-.18em R}} \font\cmss=cmss10 
\font\cmsss=cmss10 at 7pt \def\IZ{\relax\ifmmode\mathchoice 
{\hbox{\cmss Z\kern-.4em Z}}{\hbox{\cmss Z\kern-.4em Z}} 
{\lower.9pt\hbox{\cmsss Z\kern-.4em Z}} {\lower1.2pt\hbox{
\cmsss Z\kern-.4em Z}}
\else{\cmss Z\kern-.4em Z}\fi} 

\overfullrule=0mm
\def\file#1{#1}
\newcount\figno \figno=0
\newcount\figtotno      
\figtotno=0
\newdimen\captionindent 
\captionindent=1cm 
\def\figbox#1#2{\epsfxsize=#1\vcenter{
\epsfbox{\file{#2}}}} 
\newcount\figno
\figno=0
\def\fig#1#2#3{ \par\begingroup\parindent=0pt
\leftskip=1cm\rightskip=1cm\parindent =0pt 
\baselineskip=11pt
\global\advance\figno by 1
\midinsert
\epsfxsize=#3
\centerline{\epsfbox{#2}}
\vskip 12pt
{\bf Fig. \the\figno:} #1\par
\endinsert\endgroup\par
}
\def\figlabel#1{\xdef#1{\the\figno}} 
\def\encadremath#1{\vbox{\hrule\hbox{\vrule\kern8pt 
\vbox{\kern8pt \hbox{$\displaystyle #1$}\kern8pt} 
\kern8pt\vrule}\hrule}} \def\enca#1{\vbox{\hrule\hbox{
\vrule\kern8pt\vbox{\kern8pt \hbox{$\displaystyle #1$}
\kern8pt} \kern8pt\vrule}\hrule}}
\def\tvi{\vrule height 12pt depth 6pt width 0pt} 
\def\tv{\tvi\vrule}

\def\IR{\relax{\rm I\kern-.18em R}}
\font\cmss=cmss10 \font\cmsss=cmss10 at 7pt 
\def\IZ{\relax\ifmmode\mathchoice
{\hbox{\cmss Z\kern-.4em Z}}{\hbox{\cmss Z\kern-.4em Z}} 
{\lower.9pt\hbox{\cmsss Z\kern-.4em Z}}
{\lower1.2pt\hbox{\cmsss Z\kern-.4em Z}} 
\else{\cmss Z\kern-.4em Z}\fi} \def\buildrel#1\under#2{ 
\mathrel{\mathop{\kern0pt #2}\limits_{#1}}}

\Title{UNC-CH-MATH-98/1}
{{\vbox {
\centerline{Folding the Square-Diagonal Lattice}}}}
\bigskip
\centerline{P. Di Francesco\footnote*{e-mail: philippe@math.unc.edu},}
\bigskip
\centerline{\it Department of Mathematics,} 
\centerline{\it University of North Carolina at Chapel Hill,} 
\centerline{\it  CHAPEL HILL, N.C. 27599-3250, U.S.A.} 
\vskip .5in
\noindent We study the problem of "phantom" folding of the two-dimensional 
square lattice, in which the edges and diagonals of each face can be folded. 
The non-vanishing thermodynamic folding entropy per face $s\simeq.2299(1)$ 
is estimated both analytically and numerically, by successively mapping 
the model onto a dense loop model, a spin model and a new 28 Vertex, 
4-color model. 
Higher dimensional generalizations are investigated, as well as
other foldable lattices.

\Date{02/98 $\ \ $ PACS: 64.60.-i $\ \ \ \ \ $
}  


\nref\KN{Y. Kantor and D.R. Nelson, {\it Crumpling Transition 
in Polymerized Membranes}, Phys. Rev. Lett. {\bf 58} (1987) 2774
and {\it Phase Transitions in Flexible Polymeric Surfaces}, 
Phys. Rev.  {\bf A 36} (1987) 4020.}
\nref\NP{D.R. Nelson and L. Peliti, {\it Fluctuations in Membranes
with Crystalline and Hexatic Order}, J. Physique {\bf 48} (1987) 1085.}
\nref\PKM{M. Paczuski, M. Kardar and D.R. Nelson, {\it Landau
Theory of the Crumpling Transition}, Phys. Rev. Lett. 
{\bf 60} (1988) 2638.}
\nref\DG{F. David and E. Guitter, {\it Crumpling Transition
in Elastic Membranes: Renormalization Group Treatment},
Europhys. Lett. {\bf 5} (1988) 709.}
\nref\BESP{M. Baig, D. Espriu and J. Wheater, {\it Phase Transitions in 
Random Surfaces}, Nucl. Phys. {\bf B314} (1989) 587;
R. Renken and J. Kogut, {\it Scaling Behavior at the Crumpling Transition},
Nucl. Phys. {\bf B342} (1990) 753; R. Harnish 
and J. Wheater, {\it The Crumpling Transition of Crystalline Random Surfaces},
Nucl. Phys. {\bf B350} (1991) 861; J. Wheater and P. Stephenson,
{\it On the Crumpling Transition in Crystalline Random Surfaces},
Phys. Lett. {\bf B302} (1993) 447.}
\nref\MEA{P. Di Francesco, O. Golinelli and E. Guitter, 
{\it Meander, Folding
and Arch Configurations}, Mathl. Comput. Modelling, 
Vol. {\bf 26}, No.8-10 (1997) 97-147, 
{\it Meanders and the Temperley-Lieb Algebra},
Commun. Math. Phys. {\bf 186} (1997), 1-59
and {\it Meanders: a direct enumeration approach},
Nucl. Phys. {\bf B482[FS]} (1996), 497-535; P. Di Francesco, 
{\it Meander Determinants},
to appear in Commun. Math. Phys. (1998)}
\nref\KAN{Y. Kantor and M.V. Jari\'c, Europhys. Lett. {\bf 11} (1990) 157.}
\nref\DGE{P. Di Francesco and E. Guitter {\it Entropy of Folding of 
the Triangular Lattice}, Europhys. Lett. {\bf 26} (1994) 455.}
\nref\DGT{P. Di Francesco and E. Guitter {\it Folding Transition of the
Triangular Lattice}, Phys. Rev. {\bf E50} (1994) 4418-4426.}
\nref\DGGF{M. Bowick, P. Di Francesco, O. Golinelli and E. Guitter 
{\it 3D Folding of the triangular lattice}, Nucl. Phys. 
{\bf B450[FS]} (1995) 463-494.}
\nref\TLA{H. Temperley and E. Lieb, {\it Relations between the Percolation
and Coloring Problems and other Graph-Theoretical Problems associated with
regular Planar Lattices: Some Exact Results for the Percolation
Problem}, Proc. Roy. Soc. {\bf A322} (1971) 251-280; see also the book
by P. Martin, {\it Potts Models and Related Problems in Statistical
Mechanics}, World Scientific, Singapore (1991) for a review.}
\nref\BAX{R.J. Baxter, {\it Exactly Solved Models in Statistical
Mechanics}, Academic Press, London (1982).}
\nref\DEG{P. Di Francesco, B. Eynard and E. Guitter, {\it Coloring
Random Triangulations}, cond-mat/9711050, to appear in 
Nucl. Phys. {\bf B} (1998).}

\newsec{Introduction}
\par
Models for polymerized membranes can help our understanding of
biological systems. 
A typical discretized model for a membrane consists of a network
of vertices (atoms) linked by bonds. Irregular networks correspond
to fluid membranes, with arbitrary connectivity at each vertex.
Regular networks are called tethered membranes: their bonds
may have short variations in length, leading to a geometrical
crumpling transition \KN. Continuous versions of these models 
have confirmed this result, both analytically \NP-\DG\ 
and numerically \BESP.
In the present paper, we consider a
discrete model for rigid bond-membranes, represented by 
regular 2-dimensional networks whose vertices are linked 
through rigid bonds of fixed length.
The only possibility for such a membrane to modidy its
spatial configuration is through folding along its bonds,
serving as hedges between adjacent faces.  
The effects of self-avoidance on discrete folding models
can be extremely complex: already in one dimension, this has
lead to interesting developments, in relation with the "meander" problem
\MEA.
The type of folding we consider however is not self-avoiding, in the
sense that we allow the membrane to interpenetrate itself
(phantom folding). 

This work follows a previous study of the folding of the 
triangular lattice \KAN, leading to an exact result for the
folding entropy in two dimensions \DGE, and to evidence
for a first order folding transition between a flat and a folded phase
\DGT, and some further developments
in which the triangular lattice is folded into the 3-dimensional
Face Centered Cubic lattice \DGGF. 

In this paper, we consider the folding problem of the 
square-diagonal lattice (see Fig.1 below), made of the square lattice
with bonds joining all first and half of second-neighbor vertices. 
In a first
step, we study the two-dimensional folding of the lattice
and obtain estimates for its folding entropy. In a second step, we
introduce $d$-dimensional generalizations in which the
lattice is folded onto a regular $d$-dimensional lattice, allowing
only for a finite number of possible relative foldings of
adjacent faces of the membrane in the target $d$-dimensional
space. Estimates for the higher-dimensional
folding entropies are also found.

\medskip
The paper is organized as follows.
In Sect.2, we introduce the 2-dimensional folding problem of the 
square-diagonal lattice as an edge tangent-vector model. A reformulation
as a colored loop model leads to some analytic bounds on the
folding entropy. The structure of the colored loop model is further
investigated in Sect.3, in relation with the Temperley-Lieb algebra
and the Potts and 6 Vertex models. This leads to better analytic 
bounds on the entropy. 
In Sect.4, we transform the colored loop model into a $28$ Vertex
model, allowing for the numerical study of the folding entropy,
carried out in Sect.5.

The next two sections are devoted to higher dimensional generalizations
of the square-diagonal lattice folding. The idea is to fold the
lattice into a target $d$-dimensional lattice. We find two such lattices
compatible with the square-diagonal lattice: the Hypercubic-Diagonal (HCD)
and Face-Centered Hypercubic (FCH) lattices. 
The HCD model is studied in Sect.6, and successively mapped onto a
colored loop model and a vertex model. Various estimates of the
folding entropy follow.
In Sect.7, an analogous study is carried out for the FCH model.
The equivalent vertex model is particularly simple and gives access
to very good numerical estimates of the folding entropy.

In Sect.8, we present a classification of all possible compactly
foldable lattices in two dimensions.
In addition to the known square and triangular lattices, we find
only two more: the square-diagonal lattice studied in this paper, and 
the double-triangular lattice, a decoration of the
triangular lattice obtained by adding vertices in the middle of
one third of its edges (one per triangle), 
and by drawing the corresponding heights. 
The latter lattice is then folded in both 2 and higher dimensions,
giving rise to new vertex models on the Kagom\'e lattice.

We gather a few concluding remarks in Sect.9.
\par
\newsec{The Folding Problem}
\par
\subsec{Folding of the Square-Diagonal lattice}
\fig{The Square-Diagonal lattice. It has two types of vertices,
respectively 4- and 8-valent, and two types of edges, short
(length $1$) and long (length $\sqrt{2}$).}{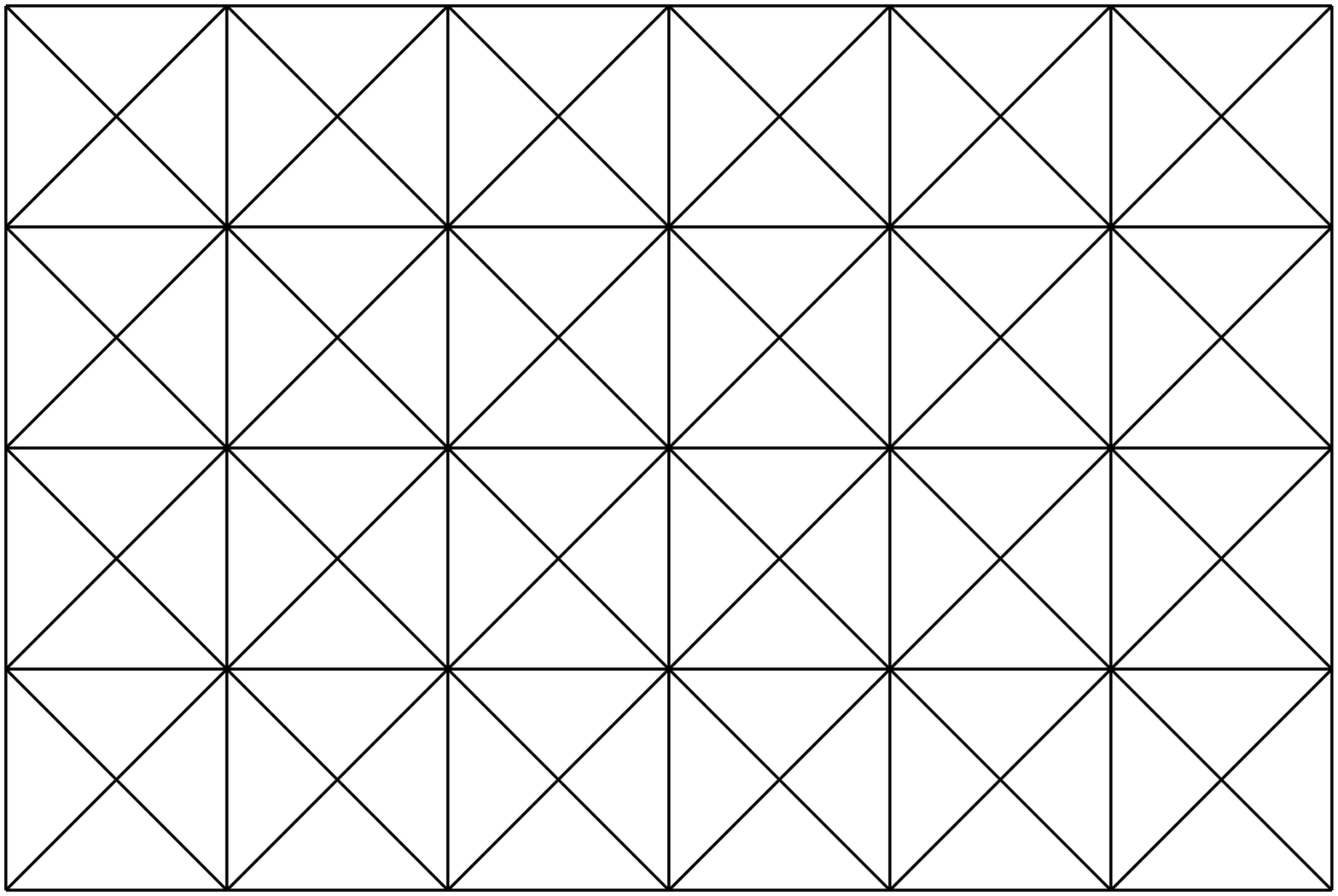}{6.cm}
\figlabel\sqdiaglat
\par
We consider the {\it Square-Diagonal lattice}, obtained from 
the standard square lattice by drawing the two diagonals 
on each face (see Fig.\sqdiaglat). 
This introduces two types of vertices, with respectively
4 and 8 incident edges, and two types of edges, short with length $1$
and long with length $\sqrt{2}$.
All the faces of the lattice are triangular, with one 4-vertex and two
8-vertices, and one long and two short edges forming a right angle.

The similarity of all the faces makes it possible to define a 
{\it complete folding} of the lattice, in which the edges serve as 
hinges between adjacent triangles, and might be either 
completely folded or not folded at all.  A {\it folding configuration} 
of the lattice is therefore a continuous map $\rho$ from the lattice to 
itself, which
preserves its faces, namely the distances between the vertices
around each triangular face.
Note that a folding configuration does not distinguish between the
various physical realizations of the actual folding of the lattice,
nor is such a configuration granted to be realizable physically.
This is called {\it phantom} folding, where the lattice is allowed 
to interpenetrate itself for the foldings to be realized, 
as opposed to the more realistic, but much more constrained 
{\it self-avoiding} folding.

\fig{The edge vectors for the square-diagonal lattice. In the 
basis $(\vec{e_1},\vec{e_2})$, the short edge vectors are of the
form $(\pm 1,0),(0,\pm 1)$, and the long edge vectors are of the
form $(\pm 1,\pm 1)$.}{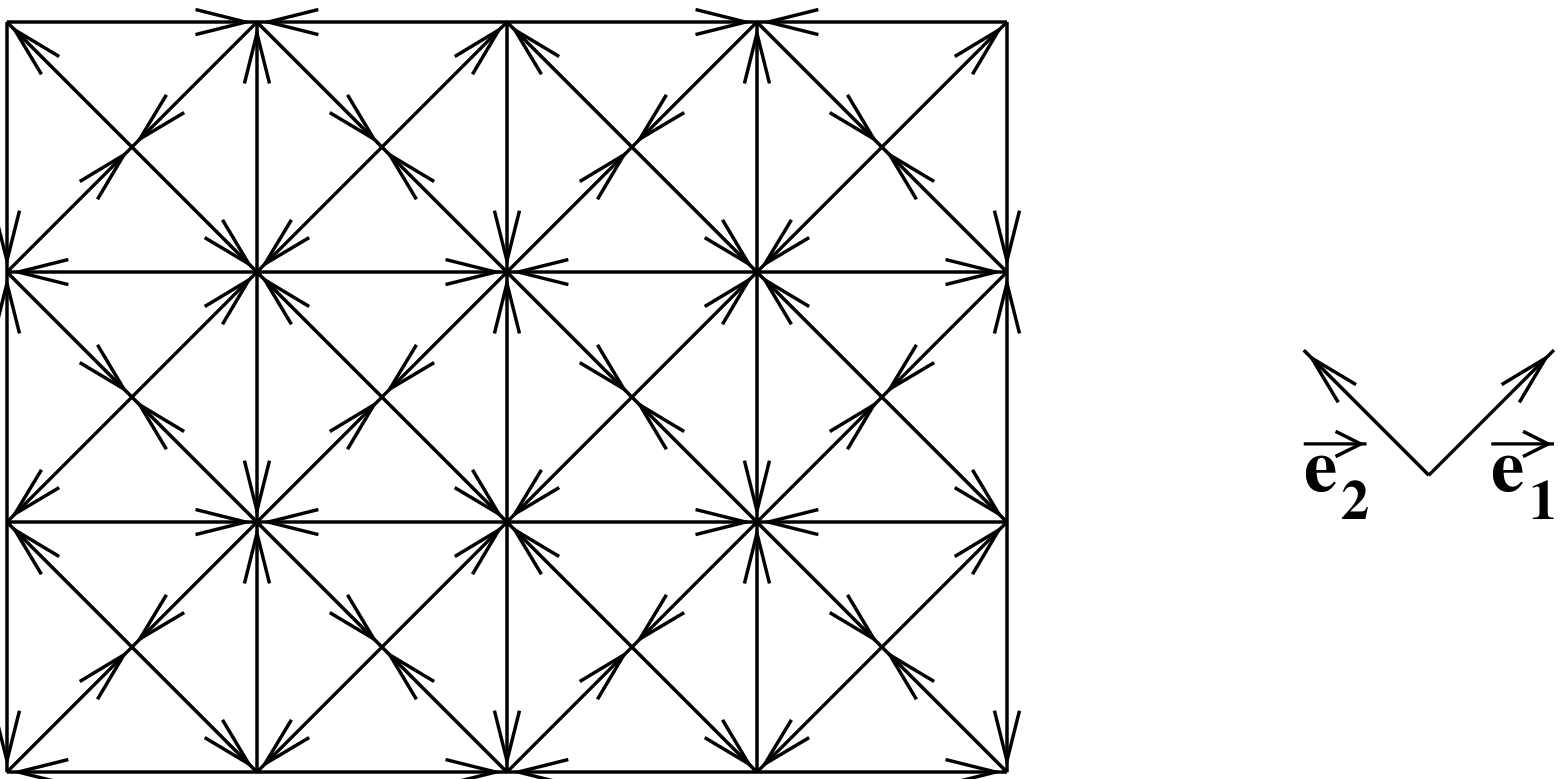}{6.cm}
\figlabel\fixor

To describe the folding configurations of our lattice, 
we note that
such a configuration is entirely determined by the list of 
all the images of the edges of the initial lattice.
Let us characterize these edges by the associated tangent
vectors $\vec{t}$, subject to the face rule
\eqn\facerul{ \sum_{\vec{t}\ {\rm around} \atop
{\rm a}\ {\rm face} }   \vec{t} ~=~ \vec{0} }
There are basically two choices of orientation of all the edge vectors,
we fix one as in Fig.\fixor.
Fixing an orthogonal basis of the plane with two short vectors, denoted
$\vec{e_1}$, $\vec{e_2}$, we see that the short edge vectors may 
only take the 4 values $\pm \vec{e_1}$ and $\pm\vec{e_2}$, whereas
the long edge vectors may only take either of the 4 values 
$\pm \vec{e_1}\pm\vec{e_2}$.

A folding configuration is characterized by the images of these tangent
edge vectors (note that short vectors are mapped to short vectors,
and long vectors to long vectors). The requirement that the faces be
preserved amounts to the condition that the face rule \facerul\ be
preserved by the map $\rho$. 
In other words,
\eqn\rulfac{ \sum_{\vec{t}\ {\rm around} \atop
{\rm a}\ {\rm face} }   \rho(\vec{t}) ~=~ \vec{0} }
This condition has to be satisfied by the images 
of the edge vectors around all the faces of the lattice.

With these conditions, the partition function $Z_{SD}$ of the
folding problem of the square-diagonal lattice (actually of a portion
thereof, made of $N$ triangular faces) is simply the number 
of distinct folding configurations, namely of distinct configurations
of edge vectors satisfying the conditions \rulfac.
The thermodynamic entropy per triangle $s_{SD}$ is then defined as the limit
\eqn\tementro{ s_{SD}~=~\lim_{N\to \infty} \ {1\over N} {\rm Log}\, Z_{SD} }

\fig{The short edge-vector images for the flat (a) 
and completely folded (b) configurations
of the square-diagonal lattice. In case (b), the whole lattice is
folded onto the shaded triangle.}{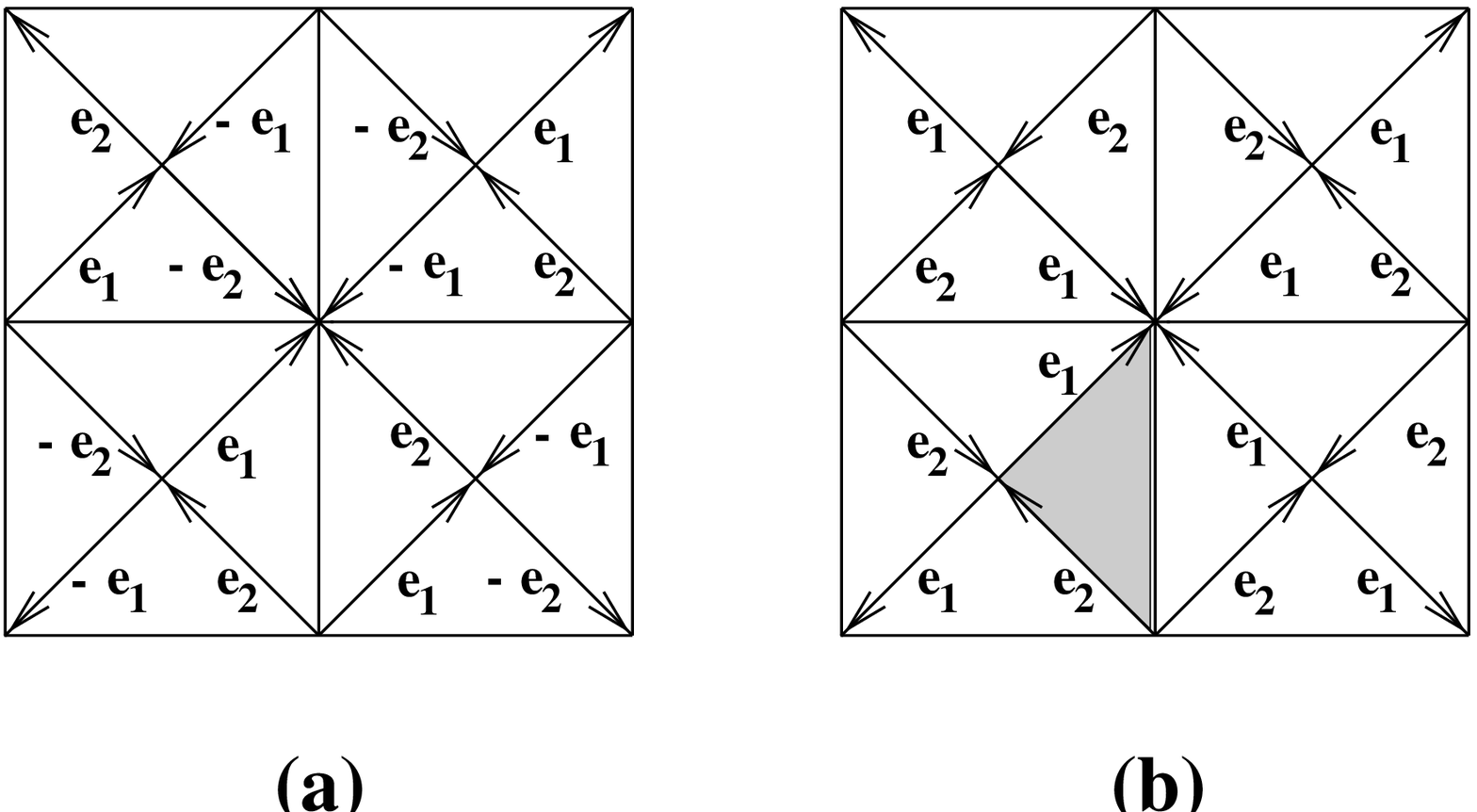}{7.cm}
\figlabel\compfol

With these definitions, we may already distinguish two particular
folding configurations: the flat configuration corresponds to no
folded edge, hence has the edge vectors of Fig.\compfol (a). 
The completely folded configuration corresponds to folding the 
whole lattice onto one single of its triangular faces. The map therefore
sends all the long edges onto one of them, say $-(\vec{e_1}+\vec{e_2})$
for definiteness, and half of the short edges to $\vec{e_1}$, the other
half to $\vec{e_2}$, as shown on Fig.\compfol (b).

\subsec{Loop Gas Reformulation}

Let us consider a folding configuration of the square-diagonal lattice.
The images of the short edge vectors characterize the configuration
completely, as the long edge vectors may be deduced from the face rules
\facerul. But these are still constrained as follows. 

\item{(i)} the two short edge vectors around each face must be 
perpendicular, i.e., one of them is equal to $\pm\vec{e_1}$ and the 
other to $\pm\vec{e_2}$.

\item{(ii)} any two adjacent triangular faces sharing a long edge 
have short edges with either of the two possible images below
\eqn\eithto{ \figbox{2.cm}{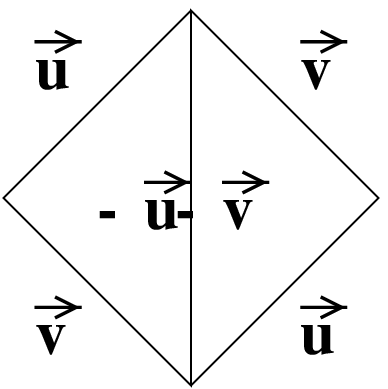} \qquad \figbox{2.cm}{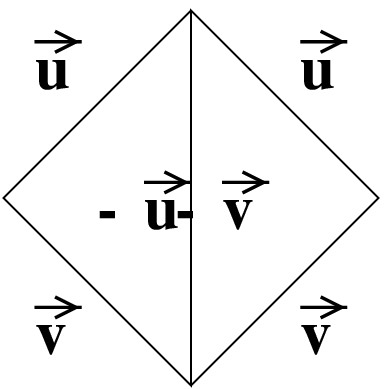}}
corresponding respectively to an unfolded or folded long edge.
\par

The two conditions (i)-(ii) above are the only constraints on the
images of the short edge vectors.
The image of a given short edge vector $\vec{t}$ reads 
\eqn\imvec{ \rho(\vec{t})~=~ \epsilon \vec{e_i}}
and is characterized by the pair $(i,\epsilon)$, where $i\in \{1,2\}$ 
can be thought of as a {\it color} of the edge,
and $\epsilon=\pm 1$ is a sign.
In the condition (ii), the signs are the same for the short edges 
of the same color. Hence these signs are preserved along chains of 
short edges throughout the lattice, forming loops of either color 1 or 2.
\fig{The dual of the square-diagonal lattice (thick lines). 
The original square-diagonal lattice is represented in thin 
lines.}{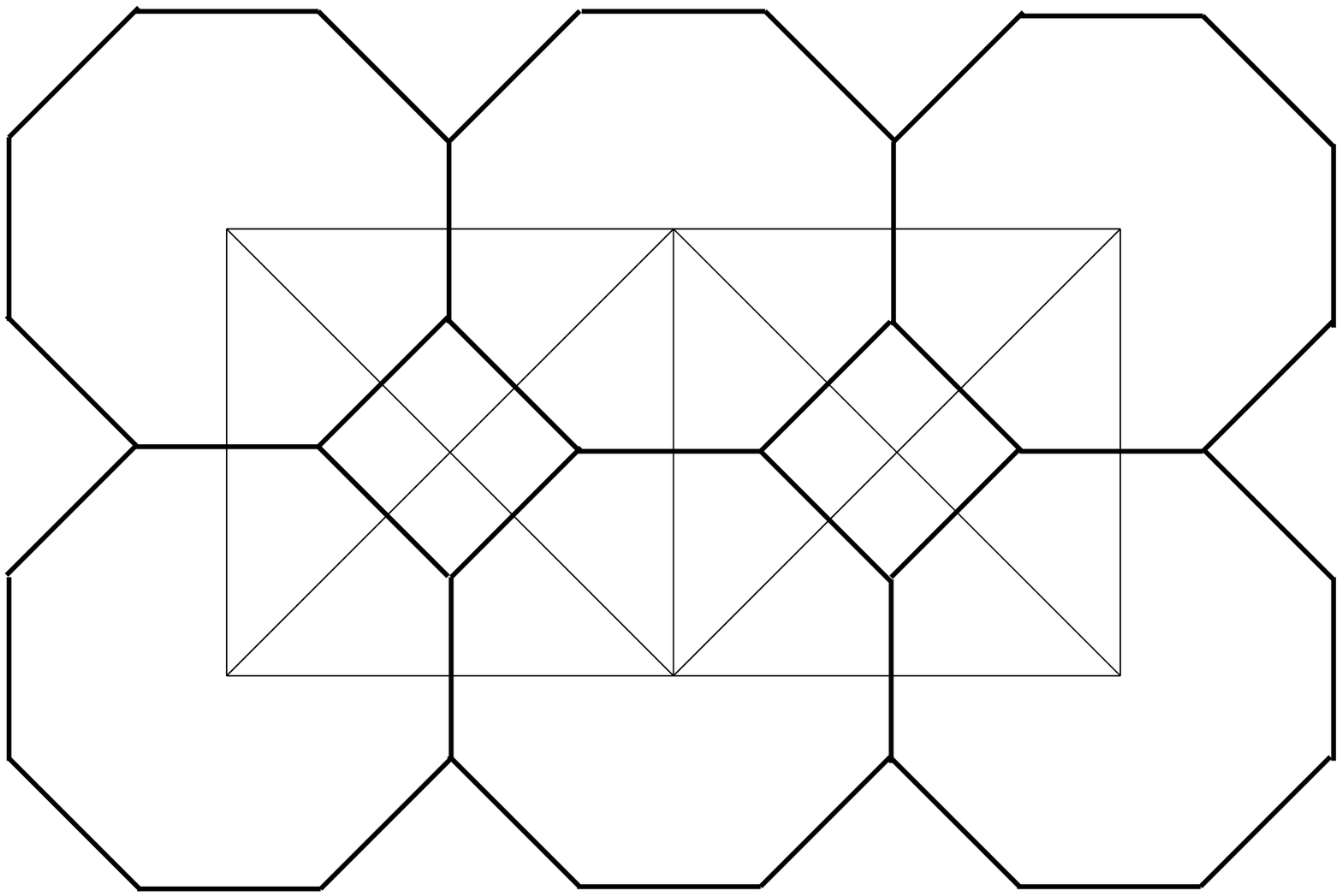}{6.cm}
\figlabel\dual
More precisely, let us consider the {\it dual} of the square-diagonal
lattice, whose vertices are the centers of the triangular faces, and
whose edges cross the former ones. This dual has only square and octagonal 
faces as depicted in Fig.\dual. The short edges are in bijective 
correspondence with the edges of the square faces in the dual lattice,
hence we paint them with the corresponding colors $i=1,2$, which we will
call black (solid lines in the pictorial representations) and white
(dashed lines in the pictorial representations).
\fig{A possible coloring of the dual of the short edges 
in black (thick solid lines) and white (thick dashed lines). The 
additional signs $\epsilon, \eta,\sigma$ of the corresponding short edge 
vectors are conserved on both sides
of each long edge (thin long lines).}{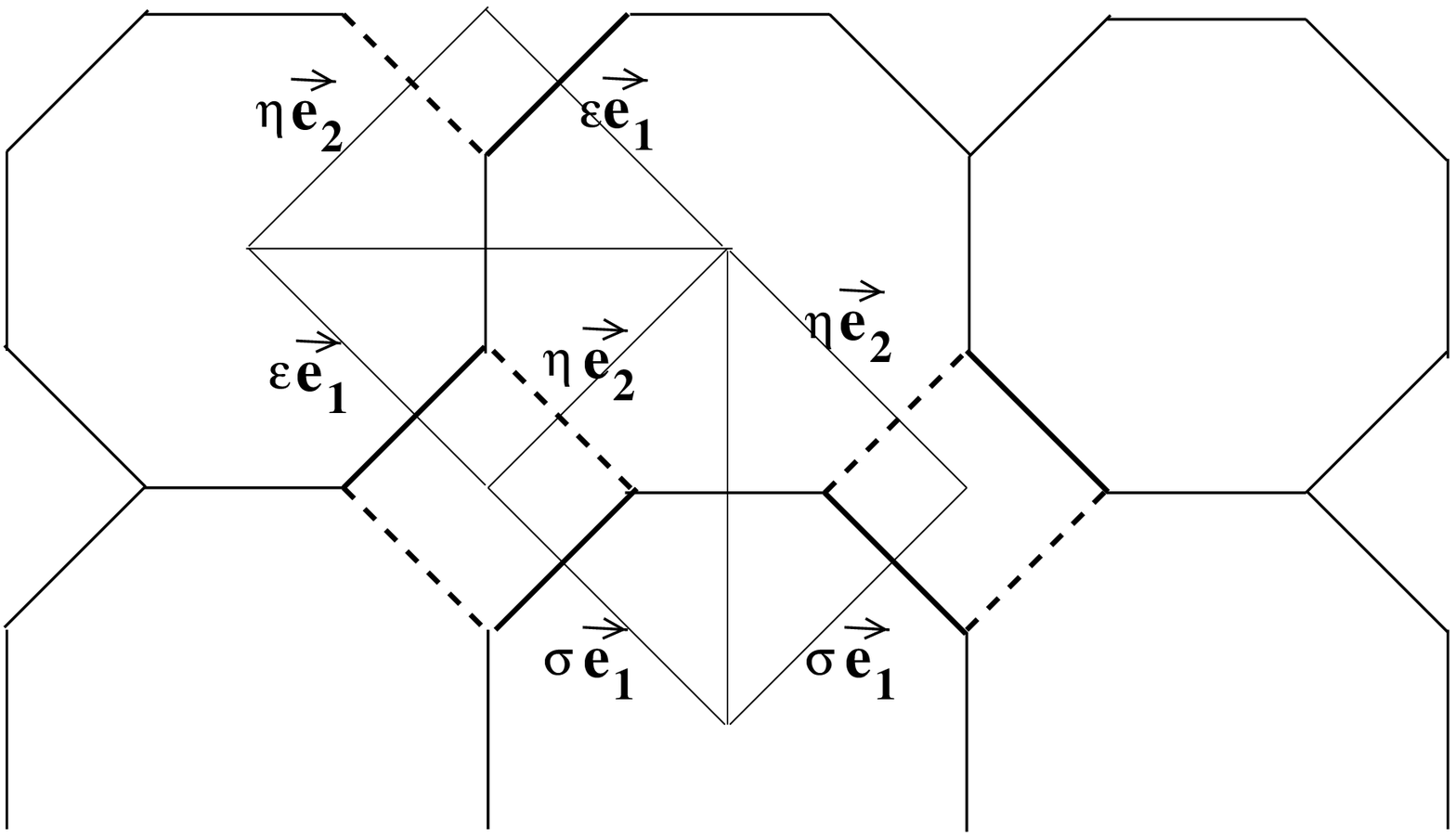}{7.cm} 
\figlabel\relaco

Around each square face of the dual, these colors must alternate, because
of the condition (i).  The two possibilities of (ii)
simply correspond to the two {\it relative} colorings of the two 
corresponding neighboring dual square faces shown in Fig.\relaco.

This suggests the following reformulation of the folding model, in the
dual picture.
We first choose an arbitrary configuration of the colorings of the
edges around the square faces. There are two choices per square face.
Such a configuration determines all the colors of all the 
short edges of the original lattice. We must now take care of the possible 
signs in \imvec. As we noted already, these signs propagate from 
short edge to short edge of the {\it same color}, between adjacent
triangles sharing a long edge.  It means that the sign remains the same
along colored loops, formed by the edges of same color, connected
in the dual lattice through a dual of long edge (namely an edge of
octagon which is not an edge of square).  

\fig{We start from an arbitrary bicoloring of the dual of the short edges,
in the dual lattice (a). We next blow up the square faces of the dual lattice 
and shrink the remaining octagon edges to zero, so as to get 
a (bicolored) square lattice (b). We have represented on (b) a new
square lattice $S$ in thin solid lines; each of its faces contains
one bicolored square.}{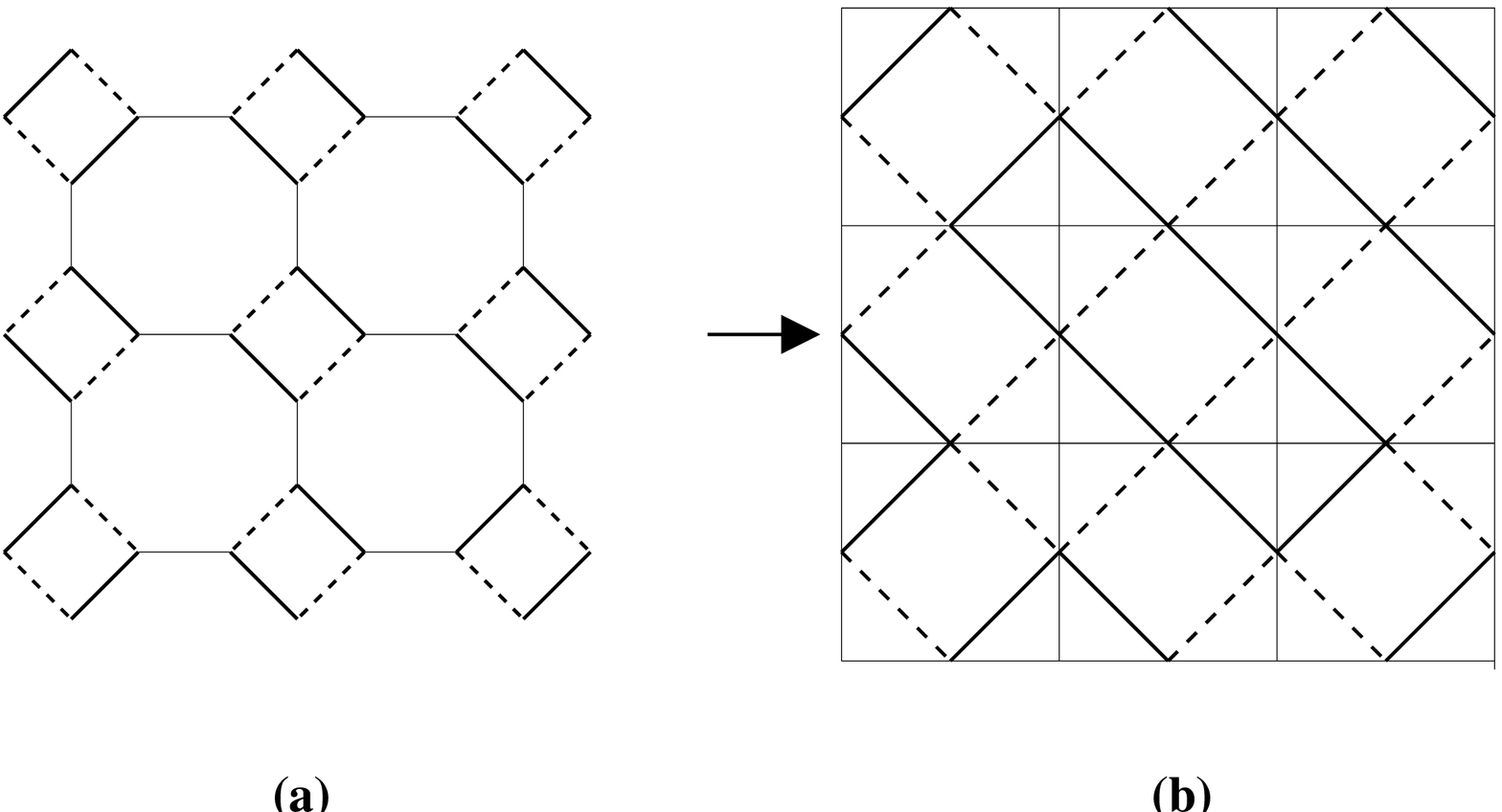}{7.cm}
\figlabel\blow

Let us now blow up the
square faces of the dual and simultaneously shrink the duals of
long edges, until the latter disappear, as shown in Fig.\blow. 
The octagonal faces become square faces, forming a checkerboard with
the square faces with colored edges.
We finally draw the square lattice (denoted by $S$) 
with vertices at the center of these new square faces (see Fig.\blow). 
The faces of this square lattice simply read either
\eqn\readfa{ \figbox{2.cm}{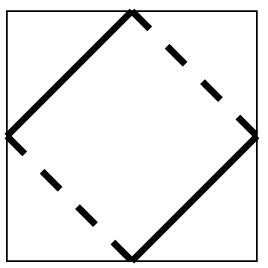}\ \ \ \  {\rm or} \ \ \ \ 
\figbox{2.cm}{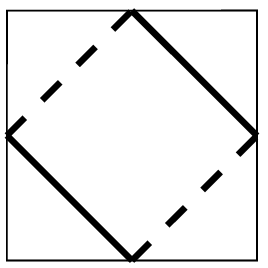} }
according to the previous coloring. Note that the black (resp. white)
edges are connected between adjacent faces, thus forming lines
across the lattice. Discarding the problem of boundary conditions,
we may assume that the lattice geometry is toroidal (namely we consider
a rectangle of size $P\times M$ of lattice, for a total
of $N=4PM$ triangular faces, with doubly periodic 
boundary conditions) and thus these lines form loops. 
The sign of \imvec\ takes the same
value along each such black or white loop. Denoting by $N_b$ (resp. $N_w$)
the numbers of black (resp. white) loops formed by the coloring configuration,
the partition function of the folding problem of the square-diagonal lattice
reads
\eqn\pfunc{ Z_{SD}~=~\sum_{{\rm coverings} \ {\rm of}\ S\atop
{\rm with}\ \figbox{.3cm}{ap.eps} \ {\rm or} \ \figbox{.3cm}{am.eps}}
2^{N_w+N_b} }
where the coverings of the faces of $S$ correspond to the various coloring
configurations, and the factor $2$ per black or white loop simply counts
the possible choices of signs in \imvec.

We have therefore reformulated the folding problem as a dense loop
gas problem, with a fugacity $2$ per loop, and the presence of 
both loops (black) and their orthogonal trajectories (white).

\subsec{Bounds and Expansions of the Folding Entropy}

As an immediate consequence of \pfunc, let us show that the square-diagonal
lattice folding problem has a non-vanishing thermodynamic entropy $s_{SD}$ 
\tementro. Indeed, we have the minoration
\eqn\minoz{ Z_{SD}~ \geq~ 2^{N/4} }
obtained by counting the coloring configurations
of the faces of S, i.e. a factor 2 per square. The factor $N/4$
follows from the fact that each square 
corresponds to 4 triangles in the original lattice. 
We therefore find 
\eqn\minentrop{s_{SD}~ \geq~ {1 \over 4}{\rm Log}\, 2~=~.1732...}
and the thermodynamic entropy does not vanish.

Another way would have consisted in choosing a particular coloring
configuration of the faces of $S$, maximizing the number of black and
white loops. This is readily obtained by letting the face configurations
alternate between the two possibilities of \readfa\ thus forming a
checkerboard.
Note that there are two such "groundstates", obtained from one another
by exchanging the two face configurations. 
Restricting to one of these two colorings of $S$, 
we get the following minoration
\eqn\minofo{ Z_{SD} \geq  2^{N_b+N_w} ~=~  2^{N/4} }
where we have counted one (black or white) loop surrounding each vertex
of $S$, hence a total number $N/4$. This again leads to \minentrop. 
 
\fig{A local excitation of a groundstate for $Z_{SD}$. It is
obtained by flipping a face of $S$, while the rest of the
groundstate configuration remains fixed. This creates 2 larger
loops of black and white colors, by suppressing two minimal
loops, one white and one black.}{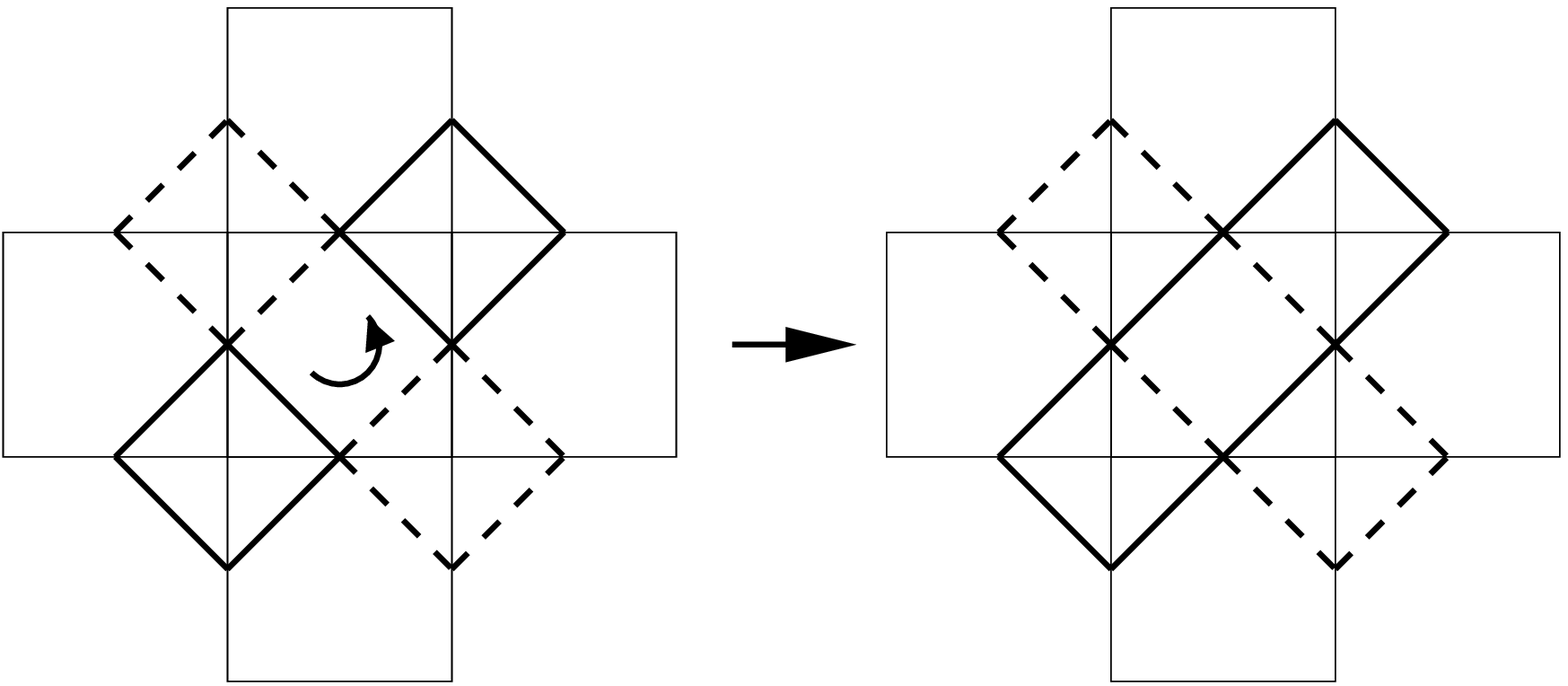}{5.cm}
\figlabel\localex

The advantage of this latter approach is to allow for a 
perturbative expansion
of $Z_{SD}$, starting from one of the above groundstates, in terms of 
elementary local excitations. Such an excitation is simply the reversal of
the coloring configuration of one face of $S$, as displayed in Fig.\localex.
Doing this will affect the 4 loops (2 white, 2 black) surrounding the face,
resulting in two larger loops, one black and one white. 
Overall this suppresses
two loops, hence contributes an extra factor 
$1/4$ to the partition function. 
There are $N/4$ of these excitations, hence
\eqn\expfun{ Z_{SD}~=~ 2^{N/4} (1+ {1 \over 4} {N\over 4} + ...)}

\fig{The four elementary excitations forming a $2\times 2$ square.
These replace 4 black and 5 white loops by 1 white and 2 black ones,
resulting in a relative weight $1/2^6$ instead of the expected $1/2^8$ for
four distant local excitations. The role of black and white loops
is exchanged when the four excitations are shifted by one 
face.}{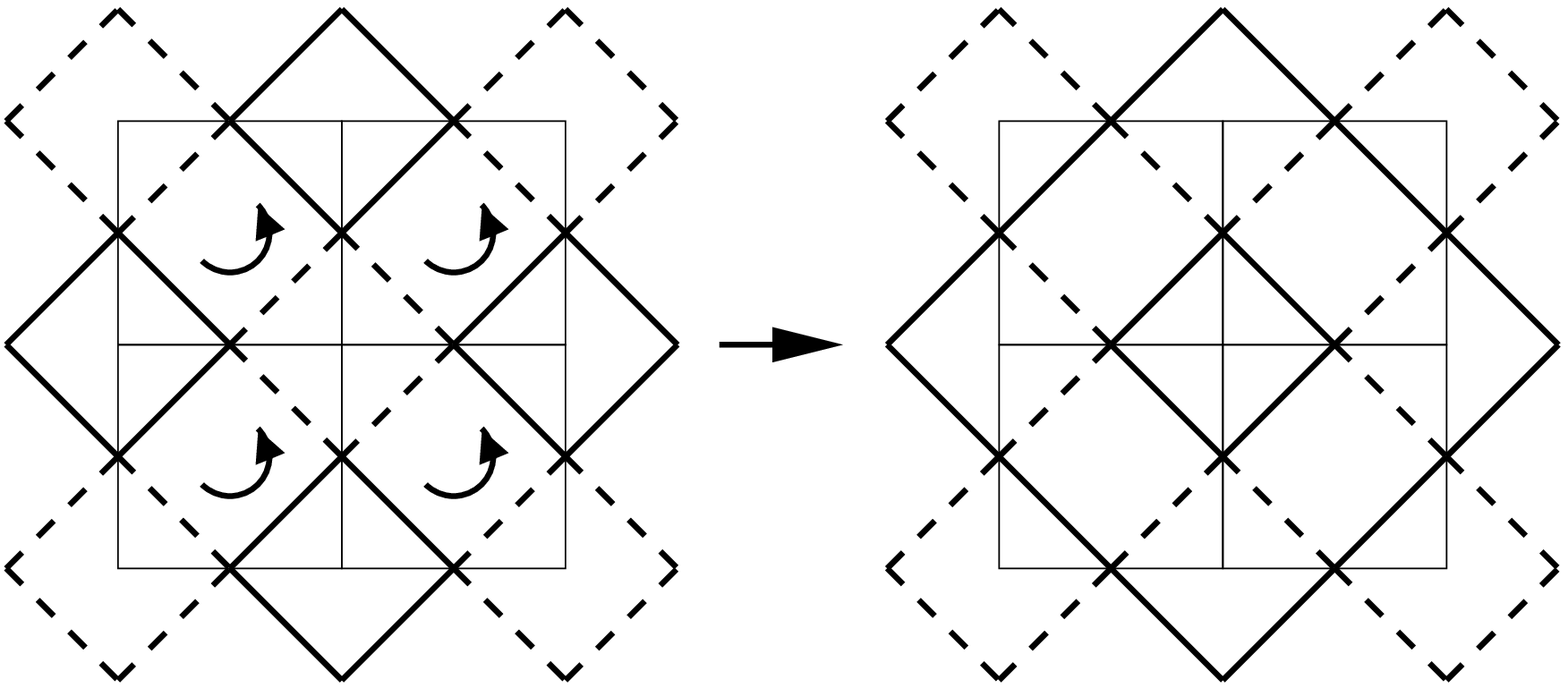}{7.cm}
\figlabel\forex

This expansion can be continued for higher numbers of excitations,
but these might interact, by creating or suppressing more loops when they
are close together than when they are separated. The first instance of this
occurs for 4 excitations.  
Generically, 4 excitations will contribute an extra factor of $1/2^8$ to
the partition function, except when these form a $2\times 2$ square,
as illustrated in Fig.\forex.
Indeed, in that case, a loop is created in the center of this square,
and the factor is increased to $4/2^8=1/2^6$.  This is interpreted as a contact
interaction energy between the excitations.
This phenomenon propagates to higher orders. For instance, this
is further increased when 6 excitations form a 
$2\times 3$ or $3\times 2$ rectangle, resulting in a factor $1/2^8$
instead of the expected $1/2^{12}$ if excitations did not interact.
Up to 6 excitations, the contributions to $Z_{SD}$ read 
(for convenience, we set $N/4=n$)
\eqn\uptofour{\eqalign{ 
Z_{SD}~&=~ 2^n \bigg( 1+ {n \over 2^2} +{n \choose 2} {1\over 2^4} 
+{n \choose 3} {1\over 2^6} +\big[ {n \choose 4}{1 \over 2^8} 
+n({1 \over 2^6}-{1\over 2^8})\big]\cr
&+\big[{n \choose 5}{1 \over 2^{10}} 
+n(n-4)({1 \over 2^{8}}-{1 \over 2^{10}}) \big]\cr
&+\big[{n \choose 6}{1\over 2^{12}}+n{(n-4)(n-5)\over 2}
({1\over 2^{10}}-{1\over 2^{12}})+2n({1\over 2^8}
-{2\over 2^{10}})\big]+ ...\bigg) \cr}} 
The first line of \uptofour\ includes terms up to four excitations.
The bracket includes the replacement of the expected $1/2^8$ term by
the actual $1/2^6$ explained above. This occurs with a degeneracy $n$,
corresponding to the freedom of moving the center of the $2\times 2$ 
excitation on the lattice. The second line in \uptofour\ corresponds
to 5 excitations. Again, we have replaced the expected $1/2^{10}$ by $1/2^8$
whenever a $2\times 2$ excitation is formed, together with another 
single excitation. Those arise with a multiplicity $n(n-4)$.
Finally, the third line of \uptofour\ corresponds to $6$ excitations.
We have first replaced the expected $1/2^{12}$ by $1/2^{10}$ whenever
a $2\times 2$ excitation is formed, together with two other 
single excitations. This occurs with the multiplicity $n\times (n-4)((n-5)/2$,
for the choice of the center of the $2\times 2$ excitation among $n$ vertices,
and the pair of remaining excitations among the remaining $(n-4)$ faces.
But doing so we have neglected the occurrence of $2\times 3$ and 
$3\times 2$ rectangular excitations, for which the previous $1/2^{10}$ 
must be replaced by $1/2^8$. This occurs $2n$ times ($n$ per
rectangular shape), but we have to subtract all the terms in which these 
rectangles have been counted as a $2\times 2$ square 
plus two excitations, namely $4n$ terms with weight $1/2^{10}$.  
We deduce the following Mayer expansion of the thermodynamic entropy
\eqn\mayer{ s_{SD}~=~ {1 \over 4}  {\rm Log}( 
2(1+{1\over 4}+{3\over 4^4}-{9 \over 4^5}+{18\over 4^6}+{1\over 4^4}+...))
=~\simeq~.231... }
(i.e. a partition function per triangle of $z_{SD}\simeq 1.259...$)
including the effects of up to 6 local excitations.
Note that there are negative and positive terms in the expansion \mayer,
so it is not clear whether this estimate lies above or below the exact 
value of $s_{SD}$.
It is possible however to prove that the first two terms give a 
strict lower bound on the entropy. This indeed amounts to neglecting
the interactions between the excitations, hence to underestimate $Z$
(as we under-count the loops).
We therefore get a first lower bound 
\eqn\lofibo{ s_{SD} ~>~ {1 \over 4} {\rm Log}({5 \over 2})~=~.22907...} 
corresponding to a partition function per triangle of $1.2574...$

\newsec{Loops and the Temperley-Lieb Algebra}

\subsec{The Dense Loop Gas}

The formulation of the square-diagonal folding problem as a coloring
problem has left us with a gas of dense black and white loops, each
weighed by a factor of 2.

Here, we recall some known facts about the dense loop gas on the 
square lattice $S$. 
The idea is to generate a dense set of black loops on $S$, in the
same way as we did for black and white loops, except that all the dashed
lines are erased. Each loop is weighed by a factor $\beta$,
resulting in a partition function
\eqn\denlo{ Z_\beta~=~ \sum_{{\rm coverings} \ {\rm of}\ S\atop
{\rm with}\ \figbox{.3cm}{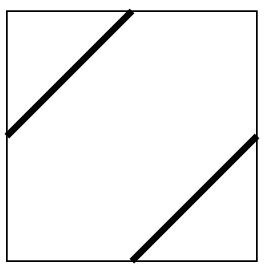} \ {\rm or} \ \figbox{.3cm}{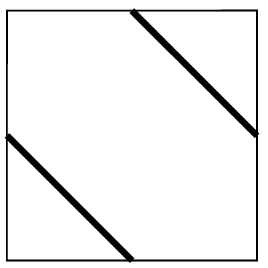}}
\beta^{L}}
where $L$ denotes the number of loops formed by the black lines.

The model is intimately related to both $Q$ states Potts
and $6$ Vertex models at infinite temperature, in that it has 
a straightforward definition in terms of the Temperley-Lieb algebra
\TLA.
Indeed, let us introduce the abstract "face" operator
\eqn\facop{\eqalign{ e_i~&=~ \figbox{5.cm}{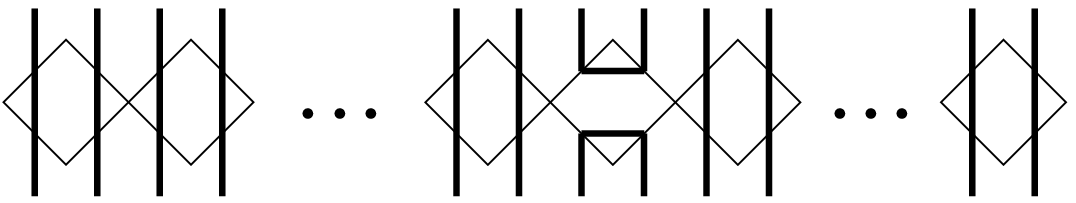}\cr
&=~ 1\otimes 1\otimes \cdots \otimes 1 \otimes e \otimes 1 
\otimes \cdots \otimes 1\cr} }
acting on a row of $2P$ parallel black lines. In \facop,
the face operator $e$ acts on the $i$-th and $(i+1)$-th lines, by connecting
them.
The definition \facop\ makes transparent the following algebraic relations
satisfied by the $e_i$'s
\eqn\tla{\eqalign{ e_i^2~&=~ \beta \ e_i \cr
e_i e_{i\pm 1} e_i ~&=~ e_i \cr
e_i e_j ~&=~ e_j e_i \ \ \ {\rm for}\ |i-j|>1 \cr}}
easily checked pictorially. The algebra generated by the abstract generators
$1,e_1,e_2,...,e_{2P-1}$, subject to \tla, 
is called the Temperley-Lieb algebra, denoted by $TL_{2P}(\beta)$. 

\fig{The first and second relation of \tla.}{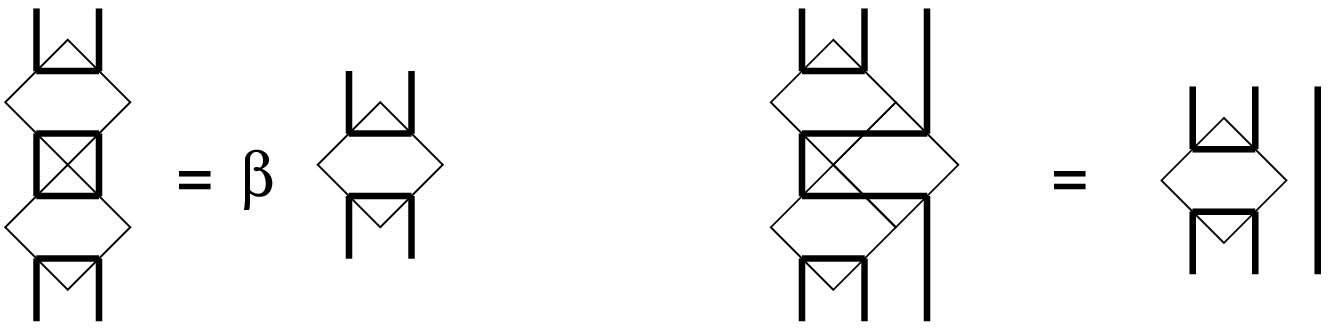}{8.cm}
\figlabel\pull

The first relation in \tla\ is consistent
with the weight $\beta$ per loop in \denlo: as shown in Fig.\pull,
we can erase the loop formed by $e_i^2$ and replace it by a factor $\beta$. 
The second relation expresses than one can "pull" the black lines,
as illustrated in Fig.\pull.
The last relation simply expresses the 
locality of the action of the face operator at lines $i$ and $i+1$.

To write the partition function of the dense loop model, we introduce
a diagonal zigzag-to-zigzag  transfer matrix
\eqn\diatr{\eqalign{ T_\beta~&=~U_\beta\ V_\beta \cr
U_\beta~&=~ \prod_{i=1}^{P-1} (1+e_{2i}) \cr
V_\beta~&=~ \prod_{i=1}^P (1+e_{2i-1}) \cr}}
The partition function of the model on a strip of width $2P$ and height $2M$,
counted in numbers of lines (with $N=4PM$, as the total number
of faces of $S$ is $N/4=PM$),
with periodic conditions along its width $2P$ boundaries reads
\eqn\periop{ Z_\beta={\rm Tr} ( T^M) }
where the trace is the standard trace on the Temperley-Lieb algebra,
defined recursively by ${\rm Tr}(1)=\beta^{2P}$  and the recursion relation
(Markov property)
\eqn\recuma{ {\rm Tr}\big(e_{i+1} E(e_1,e_2,...,e_i)\big)~=~ 
{1 \over \beta} {\rm Tr}\big(E(e_1,e_2,...,e_i)\big)}
for any expression $E$ depending on the $e_k$, $k\leq i$ only.
With this definition, \periop\ is calculated by simply first expanding
$T^M$ as a sum of products of $e$'s, then by identifying
the black lines along the width $2P$ boundaries, and replacing each black
loop by a factor of $\beta$, thus realizing exactly the sum in \denlo.
The dense loop gas partition function \periop\ coincides with
that of the square lattice isotropic $Q$-states Potts model at its critical
temperature, with $\beta=\sqrt{Q}$ \BAX.

The abstract definition \tla\ of the algebra of the $e$'s makes it 
possible to calculate \periop\ by choosing a particular representation
for the algebra. A particular choice relates it
to the partition function of the $6$ Vertex model \BAX, solved with standard
Bethe Ansatz techniques.  This gives an exact formula
for the thermodynamic entropy per site of the
dense loop model \BAX\
\eqn\termbet{s_\beta~=~ \left\{ \matrix{ 
\int_{-\infty}^{\infty} {\sinh(\pi-\mu)x \ \tanh \mu x \over
2x \ \sinh \pi x} &{\rm for} \ \ \beta=2 \cos \mu, \ \ 0<\mu<\pi\cr
{\lambda\over 2} + \sum_{n=1}^\infty 
{e^{-n\lambda}\over n}\tanh n\lambda  &{\rm for} \ \ \beta = 2 
\cosh \lambda, \ \ \  \lambda>0 \cr
2 \ {\rm Log}\, {\Gamma({1\over 4}) 
\over 2\Gamma({3 \over 4})} &{\rm for} \ \ \beta =2\cr}
\right.   }
\noindent{}This takes care of all the values of $\beta \geq 0$. 
Note that there are $N/4$ sites in the model, as there are 
4 triangles of the original square-diagonal lattice on each face of $S$.
The entropies {\it per triangle} are therefore those of \termbet\
divided by 4. 

\subsec{More Bounds on the Folding Entropy}

As the dense loop model \denlo\ is obtained by dropping the white loop 
contributions to the square-diagonal folding model \pfunc, we
get a minoration of the thermodynamic entropy \tementro\
\eqn\minotro{  s_{SD} ~>~ {1\over 4}\, s_{\beta=2}~=~
{1 \over 2}\, {\rm Log}\, {\Gamma({1\over 4})
\over 2\Gamma({3 \over 4})}~=~.19579... }
as there are 4 triangles of the original lattice per vertex of $S$.
This is below our first estimate \mayer, 
and our previous lower bound \lofibo.

We can also find an upper bound for the folding entropy $s_{SD}$, by noticing that 
the maximum possible number of white loops is obtained in one of
the two abovementioned groundstates, hence $N_w\leq N/8$ in all
configurations. Using \pfunc, we arrive at
\eqn\majop{Z_{SD}~<~2^{N/8}\, \sum_{{\rm coverings} \ {\rm of}\ S\atop
{\rm with}\ \figbox{.3cm}{ap.eps} \ {\rm or} \ \figbox{.3cm}{am.eps}}
2^{N_w}~=~ 2^{N/8} Z_2}
hence the upper bound on the folding entropy
\eqn\upfoen{ s_{SD}~< ~ {1 \over 8}\, {\rm  log}\, 2 
+ {1 \over 4} s_2~=~.28244...}
This lies above our first estimate \mayer.

This bound can be improved greatly by using the H\"older inequality
for averages, namely
\eqn\holde{ \langle A B \rangle ~\leq ~ 
\langle A^\mu \rangle^{1 \over \mu}
\langle B^\nu \rangle^{1 \over \nu} }
where $\mu,\nu$ are two positive real numbers
subject to $1/\mu+1/\nu=1$,
and $A$, $B$ are two observables averaged over a set of 
configurations $\cal C$: $\langle A \rangle =(\sum_{c\in {\cal C}}
A(c) )/|{\cal C}|$. 
We may apply \holde\ to the sum over bi-coloring configurations
of the faces of $S$, with the observables $A(c)=2^{N_b}$ and
$B(c)=2^{N_w}$. It is easy to see that the lowest upper bound 
corresponds to $\mu=\nu=2$, with the result
\eqn\resuma{ Z_{SD}~\leq ~ \sum_{{\rm coverings} \ {\rm of}\ S\atop
{\rm with}\ \figbox{.3cm}{ap.eps} \ {\rm or} \ \figbox{.3cm}{am.eps}}
2^{2 N_w}~=~Z_{\beta=4} }
where we have identified the partition function of the dense loop
model \denlo\ at $\beta=4$. 
Using \termbet, we find that $2 \cosh\, \lambda=4$,
hence $\lambda={\rm Log}(2+\sqrt{3})$, and we have the upper bound
on the folding entropy
\eqn\apentr{\eqalign{
s_{SD}~&\leq ~{1\over 4} s_{4}~=~{1 \over 8}{\rm Log}(2+\sqrt{3}) \cr
&+{1\over 4}\sum_{n=1}^\infty{1 \over n(2+\sqrt{3})^n} {(2+\sqrt{3})^{2n}-1
\over(2+\sqrt{3})^{2n}+1}~=~ .23352...\cr}}
corresponding to a partition function per triangle of $1.2630...$

Note that this upper bound only exceeds our Mayer expansion estimate by 
$1/2 \%$. 
The dense loop model at $\beta=4$ is therefore a good approximation of
the square-diagonal folding problem, as far as the entropy is concerned.
This is best seen by examining the relative numbers
of black and white loops in the sum \pfunc. 
In the groundstates, $N_b=N_w=N/8$. 
The first excitations (up to 3) preserve $N_b=N_w$, but change their value.
The first relative change of $N_b$ and $N_w$ is obtained for 4 excitations,
when they form a $2 \times 2$ square (c.f. the Mayer expansion \mayer). 
According to whether the central loop is black or white, we get $N_b=N_w\pm 2$.
However, we expect in average the numbers of black and white loops to
be sensibly the same. 
This explains why $Z_{\beta=4}$ is a good approximation of $Z_{SD}$,
which amounts to simply replacing 
the summand $2^{N_b+N_w}$ in \pfunc\ by $2^{2 N_w}=4^{N_w}$. 
 
\subsec{Black and White Loops}

The full black and white loop model \pfunc\ can be expressed in 
terms of {\it two} coupled Temperley-Lieb algebras, one for each color
of loop. Let us slightly generalize \pfunc\ into
\eqn\pfzb{ Z_{\beta,\bar \beta}~=~ \sum_{{\rm coverings} \ {\rm of}\ S\atop
{\rm with}\ \figbox{.3cm}{ap.eps} \ {\rm or} \ \figbox{.3cm}{am.eps}}
\beta^{N_b} {\bar \beta}^{N_w} } 
by affecting a weight $\beta$ per black loop and $\bar \beta$ per white loop.

Tilting the lattice $S$ by 45 degrees, we are led to
the introduction of the following face operators:
\eqn\opfac{\eqalign{ 
e_i~&=~ \figbox{7.5cm}{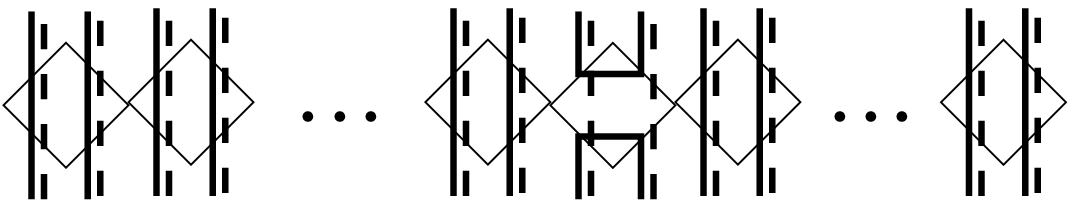} \cr
&=~ (1\otimes 1)\otimes ...\otimes (1 \otimes 1)\otimes (e \otimes 1)
\otimes (1\otimes 1)...(1 \otimes 1) \cr
f_i~&=~\figbox{7.5cm}{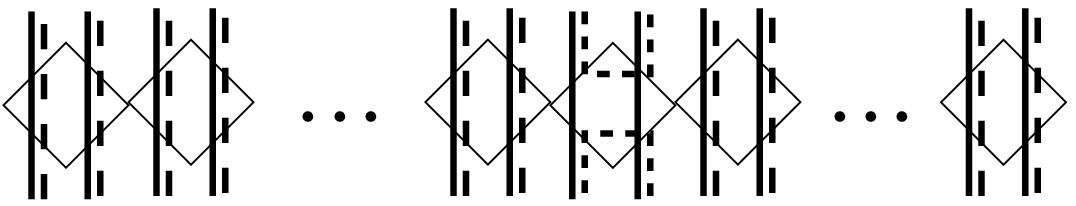} \cr
&=~ (1\otimes 1)\otimes ...\otimes (1 \otimes 1)\otimes (1 \otimes e)
\otimes (1\otimes 1)...(1 \otimes 1) \cr}}
acting on a set of $2P$ pairs of black and white parallel lines
(in each parenthesis of \opfac, the first term of the tensor product 
corresponds to the black lines and the second one to the white lines).
It is clear from the discussion of the previous section that the $e_i$
satisfy the relations \tla\ of the Temperley-Lieb algebra
$TL_{2P}(\beta)$, whereas the $f_i$ satisfy those of 
$TL_{2P}(\bar \beta)$. 
In addition, we have the commutation relations $[e_i,f_j]=0$ for
all $i$ and $j$.

Let us introduce the zigzag-to-zigzag transfer matrix
\eqn\trzf{ \eqalign{ T~&=~ U\ V \cr
U~&=~ \prod_{i=1}^P (e_{2i}+f_{2i}) \cr
V~&=~ \prod_{i=1}^P (e_{2i-1}+f_{2i-1}) \cr}}
acting on a row of $2P$ pairs of parallel black and white lines, and 
reproducing the two possible face colorings \readfa.
The partition function for a portion of size $2P \times 2M$
of the square-diagonal lattice can be finally expressed as 
\eqn\finrt{ Z_{\beta,\bar \beta}~=~ {\rm Tr}(T^M)}
by imposing periodic conditions along the width $2P$ zigzag boundaries,
namely by identifying all black and white lines along those. 
In \finrt, the trace is defined for a tensor product of any two
elements $E\in TL_{2P}(\beta)$ and $F\in TL_{2P}(\bar \beta)$ as
Tr$(E\otimes F)=$Tr$(E)\ $Tr$(F)$, and extended by linearity.

The remark of the previous section about the independence of $Z_\beta$ 
on the particular representation chosen for $e_i$ is still valid
here, and extends to the choice of representation for $f_i$ as well.
This would enable us for instance to map the model onto a pair
of coupled 6 Vertex models. We will find an equivalent 28 Vertex model
in the next section.

Comparing the expression \trzf\ for the transfer matrix to that of 
the dense loop model \diatr, we note that
\eqn\trick{\eqalign{ e_i+f_i ~&=~ {1 \over 2}\big( 
(1+e_i)(1+f_i)-(1-e_i)(1-f_i)\big) \cr
&=~ {1 \over 2} \sum_{\sigma_i=\pm 1}\sigma_i
(1+\sigma_i e_i)(1+\sigma_i f_i) \cr} }
This suggests to define a multi-parameter transfer matrix for 
the dense loop model, namely
\eqn\multt{\eqalign{
T_\beta(x_1,x_2,...,x_{2P-1})&=U_\beta(x_2,,x_4...,x_{2(P-1)}) 
V_\beta(x_1,x_3,...,x_{2P-1})\cr
U_\beta(x_2,x_4...,x_{2(P-1)})~&=~\prod_{i=1}^{P-1}(1+x_{2i} e_{2i})\cr
V_\beta(x_1,x_3,...,x_{2P-1})~&=~\prod_{i=1}^P (1+x_{2i-1}e_{2i-1})\cr}}
so that
\eqn\fint{T~=~\sum_{\sigma_i=\pm 1\atop i=1,2,...,2P-1}
\bigg(\prod_{i=1}^{2P-1}{\sigma_i \over 2}\bigg) \ 
T_\beta(\sigma_1,\sigma_2,...,\sigma_{2P-1})
T_{\bar \beta}(\sigma_1,\sigma_2,...,\sigma_{2P-1})}

The matrices $T_\beta(x)=T_\beta(x,x,...,x)$ commute with each other for 
distinct values of $x$, as a consequence of the Yang-Baxter equation \BAX.
Unfortunately, we have not been able to use this fact to "baxterize" the matrix 
$T$ by introducing a spectral parameter $x$, so as to form a 
family of commuting transfer matrices
$T(x)$. This is because the matrices $T_\beta(\sigma_1x,...,\sigma_{2P-1}x)$ 
do not
commute with each other for arbitrary values of the $\sigma_i$'s. 
So the expression \fint\ cannot be used to diagonalize $T$ in 
an efficient manner.

\newsec{An Equivalent 28 Vertex Model}

\subsec{Spin Model}

The square-diagonal folding problem in its final form can be
rewritten as a {\it spin} model as follows. To each edge
of the square lattice $S$, we attach two spin variables
$\sigma,\tau\in\{-1,1\}$. These stand for the two signs
attached to the black and white lines passing through the
center of the edge. These spins interact around each face of $S$,
because of spin conservations imposed by either of the two face 
colorings \readfa.
More precisely, each configuration of spins $(\sigma,\tau)$ around
a face of $S$ receives the Boltzmann weight
\eqn\bospi{\eqalign{ w\left(\figbox{2.cm}{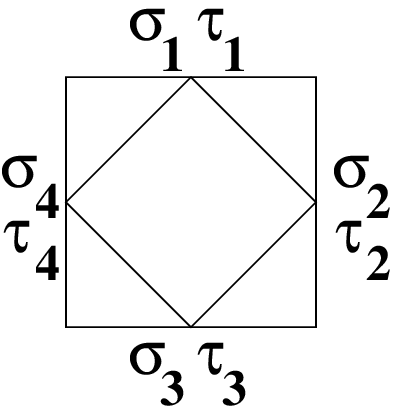}\right)~&=~
\delta_{\sigma_1,\sigma_4}\delta_{\sigma_2,\sigma_3}
\delta_{\tau_1,\tau_2} \delta_{\tau_3,\tau_4} \bigg(
\figbox{.7cm}{ap.eps} \bigg) \cr
&+\delta_{\sigma_1,\sigma_2}\delta_{\sigma_3,\sigma_4}
\delta_{\tau_1,\tau_4} \delta_{\tau_2,\tau_3} \bigg(
\figbox{.7cm}{am.eps} \bigg) \cr   
}}
where the two contributions correspond to the two possible colorings of
the face.

This formulation permits to define a row-to-row transfer matrix for the
model on a strip of width $P$, namely
\eqn\rowtran{
T_{\{(\sigma_i,\tau_i)\},\{(\sigma_i',\tau_i')\}}~=~
\sum_{s_i,t_i=\pm 1 \atop  i=1,2,..,P+1} 
\prod_{i=1}^P w\left(\figbox{2.cm}{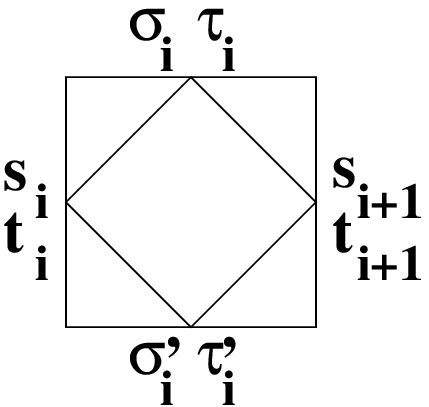}\right)}

\subsec{28 Vertex Model}

We will now rewrite the Boltzmann weight \bospi\ in terms of a 
"color" variable
$a \in \IZ_4$, defined on the edges of the dual $S^*$ of the square
lattice $S$. Namely, we replace each spin configuration around the
faces of $S$ by a coloring configuration of the edges adjacent to the
corresponding dual vertex 
\eqn\colovert{ 
\figbox{2.cm}{face.eps} \ \ \ \ \to \ \ \ \ \figbox{2.cm}{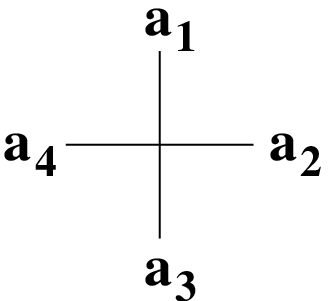} }
with the correspondence
\eqn\correspin{  \eqalign{  (\sigma,\tau) ~&\to~ a \cr
(+,+) ~&\to~ 0 \cr
(+,-) ~&\to~ 1 \cr
(-,-) ~&\to~ 2 \cr
(-,+) ~&\to~ 3 \cr}}

\fig{The 28 possible vertices corresponding to the spin
configurations of \bospi. The color $a$ may take any value
$0,1,2,3$ modulo 4. The corresponding Boltzmann weight is indicated
below each vertex.}{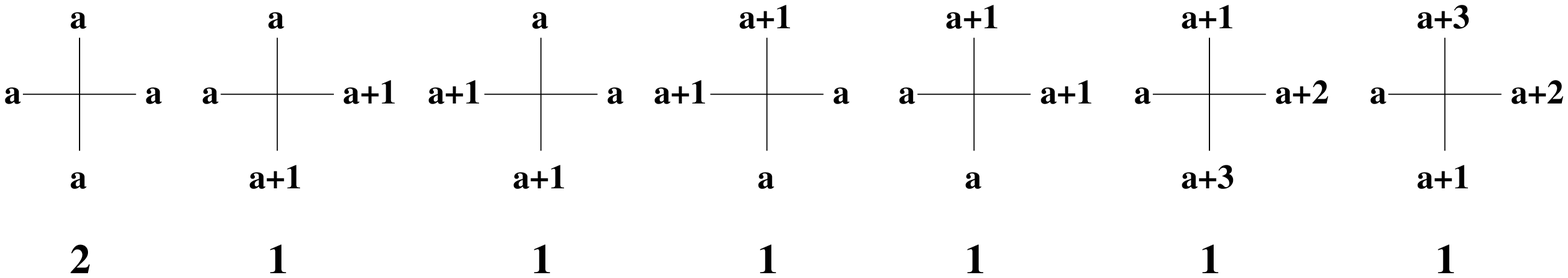}{11.cm}
\figlabel\verset

By inspection of all the possibilities in \bospi, we are left with the
only possible vertices of Fig.\verset, with $a=0,1,2,3$.
This gives a total of $4\times 7 =28$ distinct vertices.
Note that the first vertex of Fig.\verset\ 
corresponds to $\sigma_1=\sigma_2=\sigma_3=\sigma_4$ and 
$\tau_1=\tau_2=\tau_3=\tau_4$ for all $a$, hence the corresponding Boltzmann weight 
receives contributions from both terms of \bospi. 
This accounts for the Boltzmann weight $2$. All other coloring configurations
have weight $1$, as only one of the two terms in \bospi\ is selected.

The row-to-row transfer matrix \rowtran\ can be reexpressed as a
row-to-row
transfer matrix for the 28 Vertex model
\eqn\rowin{
T_{\{a_i\},\{a_i'\}}~=~
\sum_{b_i\in \IZ_4 \atop  i=1,2,..,P+1} 
\prod_{i=1}^P w\left(\figbox{2.cm}{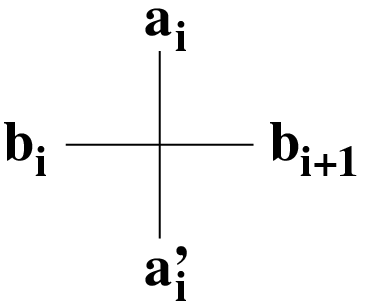}\right)}
In view of trying to solve our original folding problem, 
we tried to find an integrable structure underlying the 28 Vertex model
of Fig.\verset, allowing for distinct Boltzmann weights $w_1,w_2,..,w_7$
for each of the $7$ ($\times 4$) vertices. It appears that no
non-trivial solutions to the Yang-Baxter equation can be found here.
Moreover, our special case $w_1=2$, $w_2=...=w_7=1$ appears to be very
singular. Indeed, the $12\times 12$ R-matrix with entries
\eqn\formatri{ R_{(a,b);(a',b')}~=~ w\left( \figbox{1.3cm}{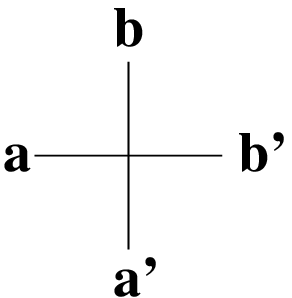}
\right) }
is not invertible at that point (note that $b=a, a\pm1$ in \formatri, 
hence the size is $12$ instead of $16$).

Although the matrix \rowin\ displays an interesting block structure, it
does not seem to be diagonalizable in a nice and systematic way.

\newsec{Numerical Studies}

This leaves us with little but the possibility of a numerical study,
which we will carry out in this section.
This is particularly efficient because $T$ is a sparse matrix. 
Indeed, there are approximately $7^N$ non-vanishing matrix elements in $T$
(given the color of the western edge of a vertex, there are
exactly 7 possibilities for the northern, eastern and southern edges),
whereas $T$ is a $4^N\times 4^N$ matrix, hence a ratio of non-vanishing
elements of $(7/16)^N$.

The results of the following subsections have been obtained by 
\item{(i)} constructing the transfer matrix $T$ of the 28 Vertex model,
including the appropriate boundary conditions
\item{(ii)} determining the block structure of $T$ 
\item{(iii)} extracting the largest eigenvalue of $T$ in the dominant block,
by iteration of the action of $T$ on an initial vector $v_0$, normalized
at each step: this process converges nicely to the Perron-Frobenius 
eigenvector of the matrix.

The particularity of the model gives us the possibility of generating
the matrix elements of $T$ "linearly", by simply listing all its 
non-vanishing entries (hence a list of length $\simeq 7^n$). 
Indeed, for each sequence of $n$ vertices in the list 
(each of which is chosen among the $7$ possibilities of Fig.\verset),
the values of the bottom and top vertical edge images are completely fixed.
Conversely, a given pair (bottom,top) of vertical edge configurations
corresponds to at most one such arrangement of $n$ vertices. This 
enables us to encode all of $T$ in a vector of length $\simeq 7^n$,
and to use it directly for determining the largest eigenvalue $\lambda_{max}$.

The various choices of boundary conditions are: 
\item{(1)} periodic, by identifying
the west-most and east-most edges of the row 
\item{(2)} fixed, by setting to 
the value $0$ both west-most and east-most edges of the row 
\item{(3)}
mixed, by fixing the east-most edge value to $0$, and summing over
all west-most edge values. This latter case is also equivalent 
to having free boundary conditions on both ends.

\subsec{Periodic Boundary Conditions}

\par\begingroup\parindent=0pt
\leftskip=1cm\rightskip=1cm\parindent =0pt
\baselineskip=11pt
$$\vbox{\font\bidon=cmr8 \def\bidondon{\bidon} \bidondon \offinterlineskip
\halign{\tv \quad # \tv 
& \hfill \ # 
& \hfill # 
& \hfill # \tv \cr
\noalign{\hrule}
\tvi  $n$ & {\rm size} &
$\lambda_{\rm max}$ \hfill & $\lambda_{\rm max}^{1/(4n)}$ \hfill \cr
\noalign{\hrule}
\tvi 1 & 1    & 2.      & 1.18920   \cr
\tvi 2 & 4    & 10.     & 1.33352   \cr
\tvi 3 & 16   & 12.9282 & 1.23773   \cr
\tvi 4 & 36   & 48.9317 & 1.27526   \cr
\tvi 5 & 256  & 83.9919 & 1.24799   \cr
\tvi 6 & 400  & 285.092 & 1.26558   \cr
\tvi 7 & 4096 & 539.435 & 1.25189   \cr
\noalign{\hrule}
}}$$
{\bf Table I:} Numerical results for the  transfer matrix
of the 28 Vertex model with periodic boundary conditions. We list the 
length $n$ of the row, the size of the relevant block of the transfer
matrix to be diagonalized, the largest eigenvalue $\lambda_{\rm max}$,
and the sequence $\lambda_{\rm max}^{1/(4n)}$, 
converging to the partition function per triangle.
Note the parity effect on this sequence, which gives a framing 
of the actual limit.
\par
\endgroup\par

We display in Table I the results for the largest eigenvalue of the
transfer matrix $T$ for periodic boundary conditions. We note the 
usual oscillatory behavior of the approximation to the partition
function per site, namely $\lambda_{\rm max}^{1/(4n)}$. 
This makes extrapolation more difficult, as we must distinguish between
both parities of $n$, and we rather rely on the other
types of boundary conditions for a better behavior.

\subsec{Fixed Boundary Conditions}

\par\begingroup\parindent=0pt
\leftskip=1cm\rightskip=1cm\parindent =0pt
\baselineskip=11pt
$$\vbox{\font\bidon=cmr8 \def\bidondon{\bidon} \bidondon \offinterlineskip
\halign{\tv \quad # \tv
& \hfill \ #
& \hfill #
& \hfill #
& \hfill # \tv \cr
\noalign{\hrule}
\tvi  $n$ & {\rm size} &
$\lambda_{\rm max}$ \hfill & $\lambda_{\rm max}^{1/(4n)}$ \hfill
& $\nu_n$ \hfill \cr
\noalign{\hrule}
\tvi 1 & 1    & 2.0000000 & 1.18920 &  \cr
\tvi 2 & 3    & 4.5615528 & 1.20889  & 1.22891 \cr
\tvi 3 & 7    & 10.898979 & 1.22025  & 1.24327 \cr
\tvi 4 & 22   & 26.562737 & 1.22748  & 1.24945 \cr
\tvi 5 & 69   & 65.399363 & 1.23247  & 1.25263 \cr
\tvi 6 & 236  & 161.98393 & 1.23612  & 1.25451 \cr
\tvi 7 & 800  & 402.76893 & 1.23890  & 1.25572 \cr
\tvi 8 & 2850 & 1004.1851 & 1.24109  & 1.25657 \cr
\noalign{\hrule}
}}$$
{\bf Table II:} Numerical results for the  transfer matrix
of the 28 Vertex model with fixed boundary conditions ($=0$ on both ends).
We list the
length $n$ of the row, the size of the relevant block of the transfer
matrix to be diagonalized, the largest eigenvalue $\lambda_{\rm max}$,
and two sequences converging to the partition function per triangle,
namely $\lambda_{\rm max}^{1/(4n)}$, and 
$\nu_n=(\lambda_{n+1}/\lambda_n)^{1/4}$.
\par
\endgroup\par

We display in Table II the results for the largest eigenvalue of the
transfer matrix $T$ for fixed boundary conditions to the value
$0$ on both ends of the row. It turns out that the block which dominates
$T$ is that containing the row-configuration $[0,0,0,...,0]$ of $n$ edges
(it gives access to the largest entry of $T$, 
namely $T_{[0...0],[0...0]}=2^n$).
The size of this block is indicated in Table II.
The sequence $\nu_n$ of consecutive ratios of eigenvalues is 
strictly increasing, and gives a good extrapolation using the Aitken
algorithm (exponential fit). We find
\eqn\extrop{ z_{SD}~=~e^{s_{SD}}~\simeq~ 1.258(1) 
\qquad \Rightarrow \qquad s_{SD}~\simeq~
.230(1) }

\subsec{Mixed Boundary Conditions}

\par\begingroup\parindent=0pt
\leftskip=1cm\rightskip=1cm\parindent =0pt
\baselineskip=11pt
$$\vbox{\font\bidon=cmr8 \def\bidondon{\bidon} \bidondon \offinterlineskip
\halign{\tv \quad # \tv
& \hfill \ #
& \hfill #
& \hfill #
& \hfill # \tv \cr
\noalign{\hrule}
\tvi  $n$ & {\rm size} &
$\lambda_{\rm max}$ \hfill & $\lambda_{\rm max}^{1/(4n)}$ \hfill
& $\nu_n$ \hfill \cr
\noalign{\hrule}
\tvi 1 & 3     & 3.0000000 & 1.316074  &  \cr
\tvi 2 & 12    & 7.7184785 & 1.291045  & 1.266492 \cr
\tvi 3 & 48    & 19.490918 & 1.280213  & 1.260593\cr
\tvi 4 & 192   & 48.985790 & 1.275350  & 1.259097 \cr
\tvi 5 & 768   & 122.95423 & 1.272000  & 1.258688 \cr
\tvi 6 & 3072  & 308.55772 & 1.269762  & 1.258630 \cr
\tvi 7 & 12288 & 774.49976 & 1.268175  & 1.258697 \cr
\tvi 8 & 49152 & 1944.7242 & 1.267000  & 1.258807 \cr
\noalign{\hrule}
}}$$
{\bf Table III:} Numerical results for the  transfer matrix
of the 28 Vertex model with mixed boundary conditions ($=0$ on one
end, free on the other). We list the
length $n$ of the row, the size of the relevant block of the transfer
matrix to be diagonalized, the largest eigenvalue $\lambda_{\rm max}$,
and two sequences converging to the partition function per site,
namely $\lambda_{\rm max}^{1/(4n)}$, 
and $\nu_n=(\lambda_{n+1}/\lambda_n)^{1/4}$. 
\par
\endgroup\par

We display in Table III the results for the largest eigenvalue of the
transfer matrix $T$ for mixed boundary conditions, fixed at $0$ on
one end, and free on the other.
Note that the size of the matrix to be diagonalized is the full size
of $T$, $3 \times 4^{n-1}$. 
On the other hand, the sequence $\nu_n$ of consecutive ratios
of eigenvalues displays a very nice convergence. 
Using the abovementioned extrapolation scheme, we arrive at
\eqn\bextrapo{ 
z_{SD}~=~e^{s_{SD}}~\simeq~ 1.2586(1) 
\qquad \Rightarrow \qquad s_{SD}~\simeq~
.2299(1) }
Note that this result is slightly smaller than the Mayer
estimate \mayer. 

\newsec{d-Dimensional Hypercubic-Diagonal Folding}

\subsec{Discrete Folding in Higher Dimensions}

In the present paper, we have studied the {\it two-dimensional} folding of 
the square-diagonal lattice, in which the image of the folding maps $\rho$ is
a subset of the original lattice. 
If we relax the latter constraint, we may just consider maps from the original
lattice to say $\IR^d$. It is however desirable that the target configurations
may only have finitely many possibilities to form local folds, i.e. we should
not allow folds with arbitrary angles. 
We may introduce a higher-dimensional {\it discrete}
folding problem by simply demanding that the image of the folding maps $\rho$, 
subject to \rulfac, be a subset of a $d$-dimensional lattice \DGGF. 
This is possible only if the $d$-dimensional "target" lattice 
is compatible with the square-diagonal lattice, 
in the sense that such $d$-dimensional folding 
configurations indeed exist. 

\fig{The unit cell of the Hypercubic-Diagonal lattice of dimension
$d=3$. The horizontal plane sections alternate between
the $A$ and $B$ types of square-diagonal lattice as
we move along the vertical. This holds for any other
plane sections perpendicular to a basis vector.}{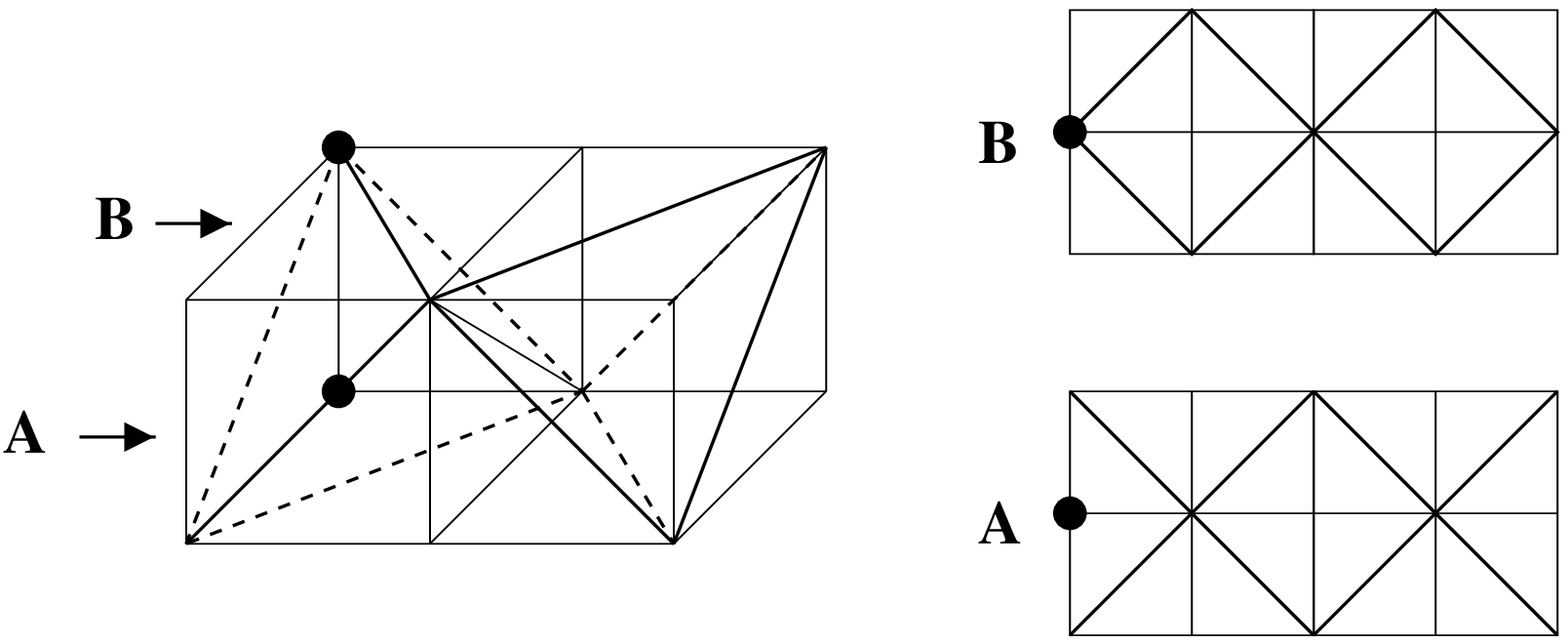}{7.cm}
\figlabel\dlat

We have found two such
compatible choices of lattice.
The first choice consists of the $d$-dimensional 
Hypercubic-Diagonal ($d$-HCD) lattice (represented in Fig.\dlat\
for $d=3$),
generated by the orthonormal vectors $\vec{e_1}$, $\vec{e_2}$, ...,
$\vec{e_d}$, $|\vec{e_i}|=1$ (with 
"short" edges of length $1$), together with exactly one of the
diagonals on each of its 2-dimensional faces (the "long" edges,
of length $\sqrt{2}$). These diagonals are chosen so that
none of them contains the origin of the lattice, and that each 
plane section of the lattice, parallel say to $(\vec{e_i},\vec{e_j})$,
$1\leq i<j\leq d$, is a copy of the square-diagonal lattice,
with an alternance of $A$ and $B$-types (c.f. Fig.\dlat) as we move
along a direction $\vec{e_k}$, perpendicular to $(\vec{e_i},\vec{e_j})$.
Note that on the lattice, any two short edges are either parallel or
perpendicular. 
The sub-lattice of the HCD formed by erasing all short edges was 
also considered for the d-dimensional folding of the triangular 
lattice in \DGGF. 

\fig{The unit cell of the $d$-dimensional Face-Centered Hypercubic 
lattice for $d=3$.}{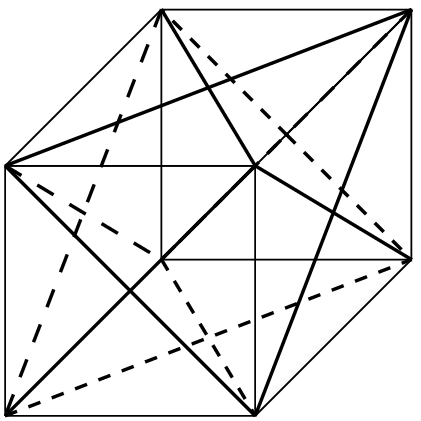}{4.cm}
\figlabel\fcc

The other is the $d$-dimensional 
Face-Centered Hypercubic (FCH) lattice (represented in Fig.\fcc\ for $d=3$),
namely the $d$-dimensional hypercubic lattice generated by 
orthogonal vectors $\vec{f_1}$, $\vec{f_2}$, ...,
$\vec{f_d}$, $|\vec{f_i}|=\sqrt{2}$ (with "long" edges
of length $\sqrt{2}$), together with the two diagonals
on each of its 2-dimensional faces, creating "short" edges of length $1$.

Both models are interesting, although it would seem at first
sight that they are dual of one another. This is not quite true, as
no transformation can exchange the roles of long and short edges: 
the short edges indeed interact around the 8 and 4-vertices 
of the square-diagonal lattice,
whereas the long edges do only around the 8-vertices.
In the remainder of this section, we will concentrate on the first case,
the $d$-dimensional HCD folding of the $2$-dimensional
square-diagonal lattice.

\subsec{The Model}

As in the 2-dimensional case, we choose
a particular "fundamental" orientation of the short and long 
edge "tangent" vectors on both the square-diagonal lattice and
the target $d$-dimensional HCD lattice, 
and consider
the set of folding configurations of the square-diagonal lattice into the
$d$-dimensional target lattice, namely the set of distinct 
images of folding maps
$\rho$ subject to the constraint \rulfac\ around each 
face of the original lattice.

The target short edges may take only $2d$ distinct values, 
namely $\pm\vec{e_1}$,
$\pm \vec{e_2}$, ..., $\pm\vec{e_d}$, where $\vec{e_1}$, ...,$\vec{e_d}$ 
denotes the orthonormal basis of the $d$-dimensional hypercubic lattice.
The target long edges may take only $2d(d-1)$ distinct values, namely 
$\pm\vec{e_i}\pm \vec{e_j}$ for $1\leq i<j \leq d$. 
The image of a given short tangent vector $\vec{t}$ of the original 
lattice reads
\eqn\dimag{ \rho(\vec{t})~=~ \epsilon \vec{e_i} }
hence is characterized by a "color" $i\in\{1,2,...,d\}$ and 
a sign $\epsilon=\pm 1$.

In a folding configuration, there are basically only two
relative possibilities for
the images of the tangent vectors to two adjacent faces
sharing a long edge. Indeed, given the
values $\vec{u}=\epsilon \vec{e_i}$, $\vec{v}=\eta \vec{e_j}$, $i\neq j$,
of the two short edges of the image of the first triangle,
there are only two possibilities 
for the images $\vec{u'}$, $\vec{v'}$
of the two short edges of the second triangle: the face rule \rulfac\
imposes that 
$\vec{u'}+\vec{v'}=\vec{u}+\vec{v}$, which holds
if and only if 
\eqn\color{\eqalign{
\vec{u'}&=\vec{u}=\epsilon \vec{e_i}\ \ \ {\rm and}\ \ \ 
\vec{v'}=\vec{v}=\eta \vec{e_j}\cr
{\rm or}\ \  \vec{u'}&=\vec{v}=\eta \vec{e_j}\ \ \ {\rm and} \ \ \ 
\vec{v'}=\vec{u}=\epsilon \vec{e_i}\cr}}
This leaves us again with the only two possibilities of \eithto.
As a corollary, the long edge adjacent to the two triangles is
always either completely folded, or unfolded.

Let us now investigate the folding configurations of short edges.
The two triangles adjacent to a short edge have images of
the form
\eqn\imsho{\figbox{3.cm}{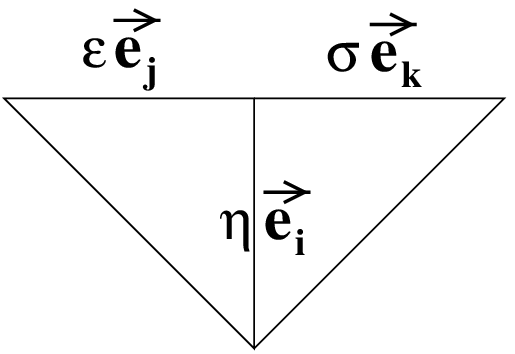} }
where $i\neq j$ and $i\neq k$.

\fig{The four types of folds of the short edge separating two 
triangles in the $d$-HCD model: complete fold, no fold, 
right angle up, right angle down.
The figure is drawn in the space spanned by 
$\vec{e_i},\vec{e_j},\vec{e_k}$ of \imsho.}{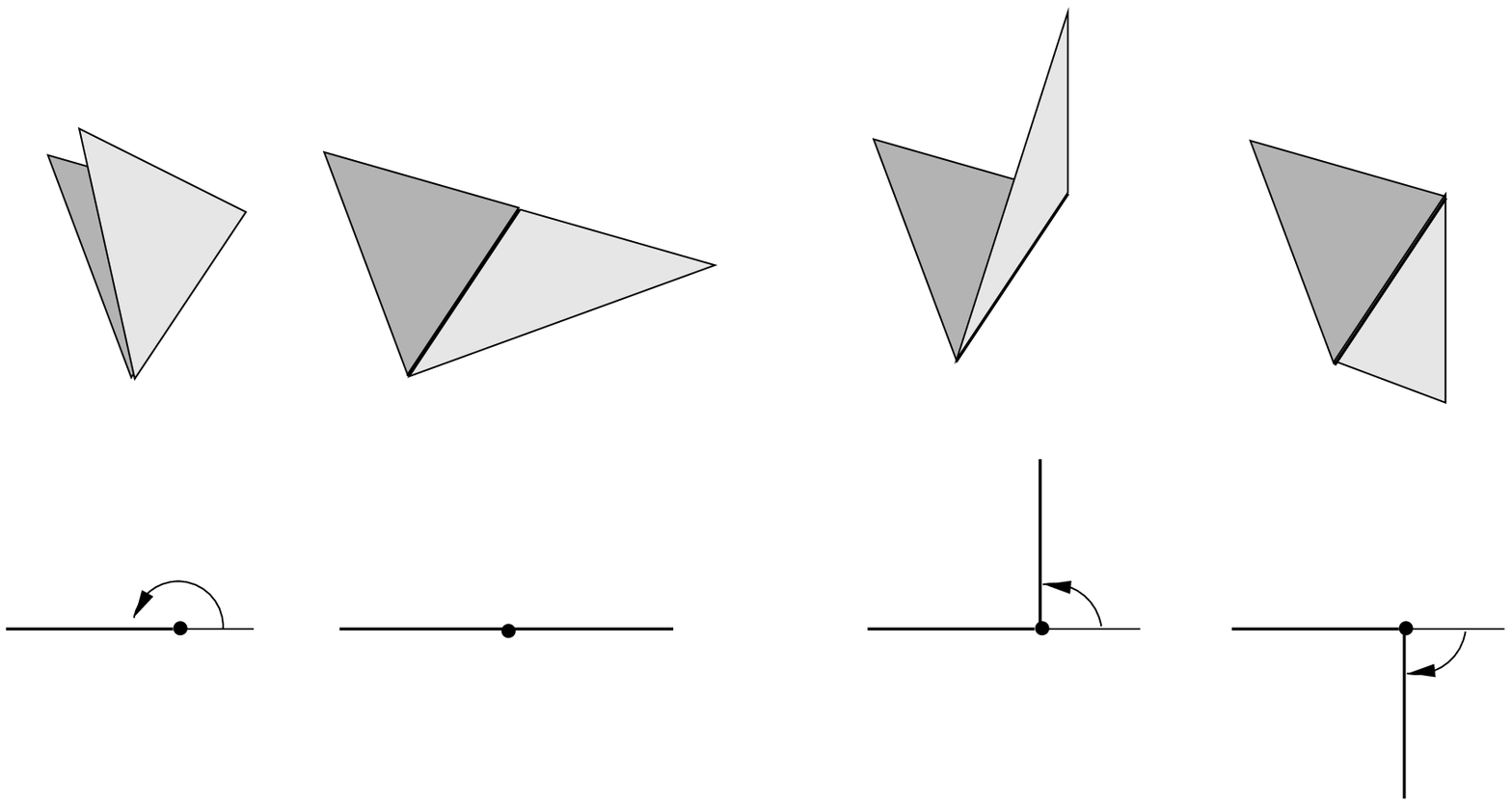}{8.cm}
\figlabel\folshort

This gives rise to essentially four types of foldings 
for the short edge, as depicted in Fig.\folshort.
If $j=k$, the short edge
is completely folded ($180^\circ$) when $\sigma=\eta$ and
unfolded when $\sigma=-\eta$. If $j\neq k$, the long edge is
always folded at a right angle, either up or down
according to the relative values of $\epsilon,\sigma$ and $\eta$.

\subsec{Equivalent Loop model}

The only constraint coming from the face rule \rulfac\ on a given triangle
is that the two short edges should have perpendicular images, of the
form $\pm \vec{e_i}$ and $\pm \vec{e_j}$ respectively, 
with $1\leq i\neq j \leq d$, 
namely that they be painted with different colors $i$ and $j$.
 
A folding configuration of the square-diagonal lattice in $d$ 
dimensions amounts to a coloring of all the short edges with colors
$i=1,2,...,d$, and a choice of signs $\epsilon=\pm 1$, which propagate
along each loop of a given color (according to \color).   
\fig{The "dual" coloring of the faces of $S$ in the $d$-HCD model.
The former short and long edges are represented in thin solid lines. The new
edges have the same colors $i$, $j$, $k$, $l$ as the short edges
they intersect.}{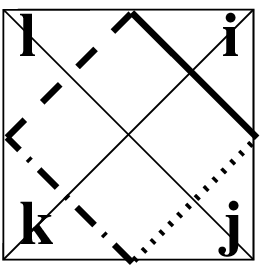}{3.cm}
\figlabel\colodu
To best see this,
let us consider the square-diagonal lattice, together with 
its square sub-lattice $S$, whose vertices are those of valency $8$
in the original lattice. 
The lattice $S$ has therefore long edges, and each of its
faces is made of $4$ former triangles, hence has 4 short edges along
its diagonals. Let us replace the coloring of the short edges by that of
"dual" edges linking the midpoints of the edges of $S$ around
its faces, as illustrated in Fig.\colodu. 
Thanks to \color, it is now clear that the new colored edges form loops of 
fixed colors, and that the signs of the tangent vectors
\dimag\ are constant along those.
Note that the coloring of the faces of $S$ must result in
the formation of loops of fixed color. This is a constraint on the coloring
of the dual edges. Namely, across each long edge separating two adjacent
faces of $S$, the colorings must be either 
reflected or propagated as follows
\eqn\proref{ \figbox{3.cm}{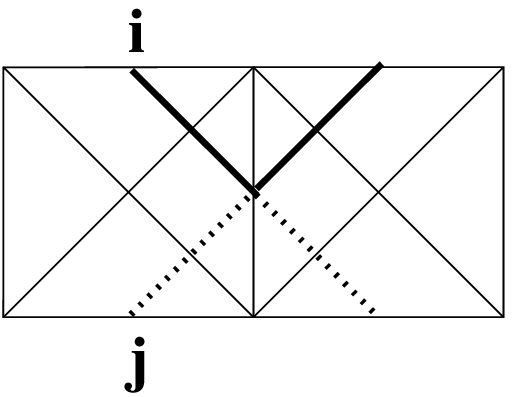} \quad {\rm or}\quad 
\figbox{3.cm}{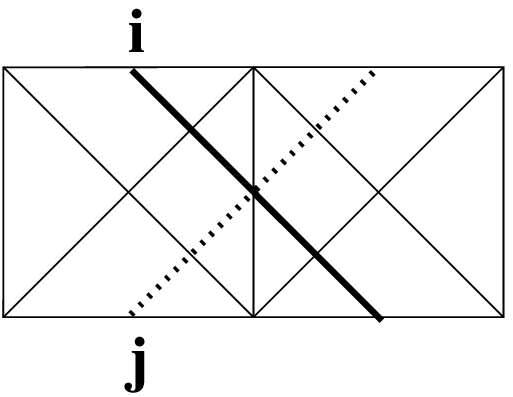} }
corresponding respectively to the two cases of \color.

\fig{A sample coloring of $S$ with $d=4$ distinct colors,
represented by solid, dashed, dot-dashed and dotted lines. The original
square-diagonal lattice is represented in thin solid lines. Note
that the lines are either propagated or reflected according to \proref\
across each long edge. The sign of the image of the short tangent
vectors is constant along each line of a given color.}{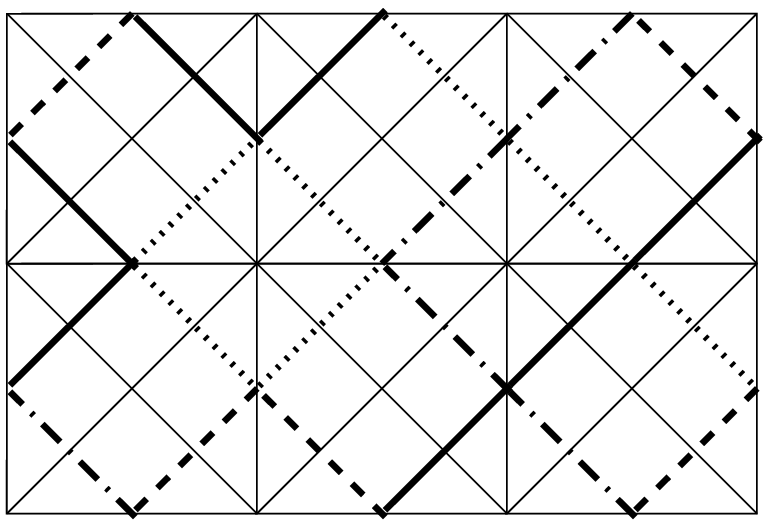}{5.cm}
\figlabel\colora

A sample coloring is displayed in Fig.\colora, with $d=4$ different colors.
Finally,
the partition function of the $d$-dimensional HCD folding of the 
square-diagonal lattice reads
\eqn\pfdcol{ Z_{HCD}^{(d)}~=~ \sum_{d-{\rm colorings} \atop 
{\rm of} \ S} 2^{N_1+N_2+...+N_d} }
where the sum extends over all possible colorings with
$d$ colors of the faces of $S$ with the gluing constraint 
\proref\ for adjacent faces. 
In \pfdcol, $N_i$ stand for the number of loops of color $i$, and the factor
$2$ per loop of a given color accounts for the two possible
choices of sign of $\rho(\vec{t})$ \dimag\ along that loop. 
Note that \pfdcol\ reduces to \pfunc\ when $d=2$.

\subsec{Estimates for the d-HCD Folding Entropy}

The expression \pfdcol\ 
for the partition function of the $d$-dimensional HCD folding
of the square-diagonal lattice provides us with upper
and lower bounds on the entropy. 

First of all, note that
\eqn\firine{ Z_{HCD}^{(d)} \leq Z_{HCD}^{(d+1)}}
as $Z_{HCD}^{(d)}$ is a partial sum of $Z_{HCD}^{(d+1)}$ corresponding to 
no loop of color $d+1$ ($N_{d+1}=0$). 
This shows that all the corresponding thermodynamic entropies 
$s_{HCD}^{(d)}$ are non-zero for $d\geq 2$. 

To find a lower bound on $s_{HCD}^{(d)}$, let us 
try to expand \pfdcol\ around one of its "fundamental" states
maximizing the total number of loops $N_1+...+N_d=N/4$. Note that there
are as many such states as d-colorings of the vertices of $S$ with colors
$1,2,...,d$ and such that no two adjacent vertices have same color. Indeed,
a fundamental state is made of loops of minimal length 4
surrounding each vertex of $S$
such that any two neighboring loops have distinct colors, hence the loop
configuration is equivalent to a vertex-coloring.
For these fundamental configurations, we have a first rough estimate
of $Z_{HCD}^{(d)}$ (actually a lower bound) in the form
\eqn\foroug{ Z_{HCD}^{(d)}~\geq~ 2^{N/4} \ Z_{S}(d) }
where $Z_S(d)$ denotes the number of $d$-colorings of the
vertices of $S$.
This is the leading order of the Mayer expansion of the partition function.
The d-coloring entropy
$s_S(d)~=~\lim_{N\to \infty} {\rm Log}(Z_S(d))/N$ is known 
exactly in the thermodynamic limit
only for $d=2$ and $3$. As we already pointed out, there
are indeed only two groundstates at $d=2$, i.e. $Z_S(2)=2$, and the 
thermodynamic entropy per site is $s_S(2)=0$: 
\foroug\ then leads to \minentrop.
For $d=3$, the $3$-coloring problem of the vertices of the square lattice $S$
(or equivalently of the faces of its dual $S^*$) is equivalent to the 
ice model \BAX, with an entropy per site
\eqn\entroba{ s_S(3)~=~ {3 \over 2} \rm Log({4\over 3}) }
This gives the lower bound for $d=3$
\eqn\lotr{ s_{HCD}^{(3)} \geq {1 \over 4}{\rm Log}\, 2+{1 \over 4} s_S(3)~=~
{\rm Log}\, 2-{3\over 8}{\rm Log}\, 3~=~.281167...}
or a partition function per triangle of $1.32467...$

Let us estimate the lower bound \foroug\ when $d$ is large.
As a very rough leading approximation, each vertex of $S$ can be
painted with one of $d$ colors, minus the small number of colors
used for its neighbouring vertices, hence if $d$ is very large, we have 
$Z_S(d)\simeq d^{N/4}$.
This gives the leading estimate 
\eqn\shcd{s_{HCD}^{(d)}~\simeq~ {1 \over 4}{\rm Log}(2d)}
A more careful study shows that this is accurate up to the order of $1/d$.

\subsec{Vertex Model}

As in the $d=2$ case, let us transform the colored loop model of Sect.6.3
into a vertex model, by mapping the coloring configurations and sign
choices on each face of $S$ onto Vertex configurations for
the dual square lattice $S^*$. Each colored new edge on the faces
of $S$ has been assigned a color $i\in\{1,2,...,d\}$ and a
sign $\epsilon\in\{-1,1\}$
(see Fig.\colodu), which are conserved (either propagated or reflected,
according to \proref)
at each crossing of an edge of $S$.
This suggests to attach to each edge of $S^*$ a {\it pair} of colors
and signs $(i,\sigma;j,\tau)$, with say $1\leq i<j \leq d$.
This gives $2d(d-1)$ possible values for this edge variable.
The vertex model is then defined by the list of all possible edge
configurations around a vertex of $S^*$, together with the gluing
property that each edge of $S^*$ has a well-defined configuration.
Each allowed vertex has Boltzmann weight $1$.

Let us count the number of allowed vertices for generic values of $d$.
This is the same as the number of coloring and sign configurations on each face
of $S$.  To count the latter, note that the only constraint on
the four color/sign pairs $(i_1,\sigma_1),...,(i_4,\sigma_4)$ is that
any two consecutive colors must be distinct. So we merely have to
count the number of cyclic arrangements of four colors taken
among $d$ around the face. Introducing the $d\times d$ matrices $I$ and $J$,
with entries $I_{i,j}=\delta_{i,j}$, and $J_{i,j}=1$ for all $i$, $j$,
the total number of vertices reads
\eqn\totver{ V_d~=~16{\rm Tr}(J-I)^4~=~16d(d-1)(d^2-3d+3)}
where we have used the relation $J^2=d J$, and factored out the contribution
$16$ for the independent choices of the four signs.

Note that \totver\ yields $V_2=32$ instead of $28$, for $d=2$.
This is because we must keep
track of the edge colors as soon as $d>2$.
At $d=2$, the model has been further
simplified by noticing that any reference to the edge coloring could be
omitted, at the expense of modifying the Boltzmann weights (four
vertices actually acquired a Boltzmann weight 2).

The row-to-row transfer matrix of the vertex model is again sparse.
Indeed, if we fix
the value of the west-most edge in a row of $N$ vertices, we have
a number of non-vanishing matrix elements of the order of
$(V_d/(2d(d-1)))^N$, whereas the matrix has a total of $(2d(d-1))^{2N}$
elements, hence a ratio $[2(d^2-3d+3)/(d^2(d-1)^2)]^N$, which
gets smaller as $d$ increases.

\subsec{Numerical Study}

In this section, we extract some numerical estimates for the
HCD folding entropy in dimension $d=3$, from the
largest eigenvalue of the transfer matrix of the 288 Vertex model
described in the previous section.

\par\begingroup\parindent=0pt
\leftskip=1cm\rightskip=1cm\parindent =0pt
\baselineskip=11pt
$$\vbox{\font\bidon=cmr8 \def\bidondon{\bidon} \bidondon \offinterlineskip
\halign{\tv \quad # \tv
& \hfill \ #
& \hfill # \tv \cr
\noalign{\hrule}
\tvi  $n$ \hfill & $\lambda_{max}$ \hfill & $\nu_n$ \hfill\cr
\noalign{\hrule}
\tvi 1 & 2.000000 & \cr
\tvi 2 & 5.587741  & 1.29286  \cr
\tvi 3 & 17.11799  & 1.32298  \cr
\tvi 4 & 54.54084  & 1.33603  \cr
\tvi 5 & 177.4631  & 1.34306  \cr
\noalign{\hrule}
}}$$
{\bf Table IV:} Numerical results for the  transfer matrix
of the $V_3=288$ Vertex model with fixed boundary conditions (
=$(1,+;2,+)$ on both ends) for the 3-dimensional HCD folding.
We have represented the size $n$ of the row, the largest eigenvalue
and the ratio $\nu_n=(\lambda_{n+1}/\lambda_n)^{1/4}$, which
converges to the partition function per triangle.
\par
\endgroup\par

We are indeed limited by the rapid growth of the size of the
transfer matrix, and more importantly of its number
of non-vanishing entries, which grows like $(8(d^2-3d+3))^n$ for a row of
width $n$. At $d=3$, this is already $24^n$, and we have been only
able to consider transfer matrices up to the width $n=5$, with fixed
boundary conditions (with the edge
variables fixed to $(i,\sigma;j,\tau)=(1,+;2,+)$ on both ends).

The results for the largest eigenvalue of the corresponding
transfer matrix of the 288 Vertex
model are displayed in Table IV, together with the sequence
$\nu_n=(\lambda_{max}(n+1)/\lambda_{max}(n))^{1/(4n)}$, converging
to the partition function per triangle $z_{HCD}^{(3)}$.
The extrapolated thermodynamic entropy reads
\eqn\expotro{ s_{HCD}^{(3)}~\simeq~ .300.. }
or a partition function per triangle $z_{HCD}^{(3)}\simeq 1.35(1)$
Note that our estimate \expotro\ agrees with the lower bound \lotr.

\newsec{d-Dimensional FCH Folding}

\subsec{The Model}

As in the 2-dimensional case, we choose
a particular "fundamental" orientation of the short and long
edge "tangent" vectors of the $d$-HCD lattice of Fig.\fcc,
and consider
the set of folding configurations of the square-diagonal lattice into the
$d$-dimensional FCH lattice, namely the set of distinct
images of folding maps
$\rho$ subject to the constraint \rulfac\ around each
face of the original lattice.

The target long edges may take only $2d$ distinct values,
namely $\pm\vec{f_1}$,
$\pm \vec{f_2}$, ..., $\pm\vec{f_d}$, where $\vec{f_1}$, ...,$\vec{f_d}$
form an
orthogonal basis of $\IR^d$ (with $|\vec{f_i}|=\sqrt{2}$ for all $i$).
The target short edges may take only $2d(d-1)$ distinct values, namely
the unit vectors
$(\pm\vec{f_i}\pm \vec{f_j})/2$ for $1\leq i<j \leq d$.

In a folding configuration, there are many possibilities for
the images of the tangent vectors to two adjacent faces
sharing a long edge, namely
\eqn\twoposd{
\figbox{3.5cm}{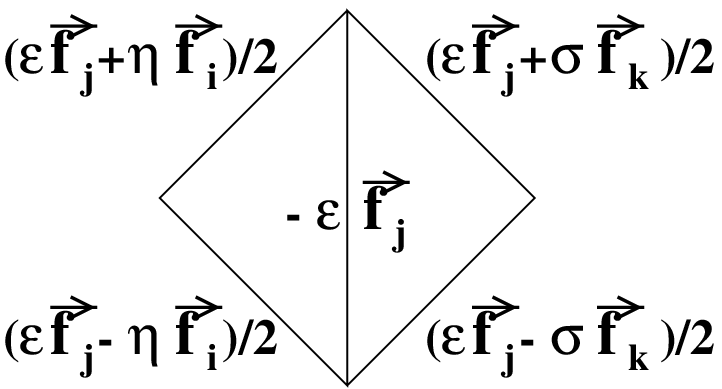} }
where $i,j,k\in\{1,2,...,d\}$,
$i$ and $k$ distinct from $j$, and
$\epsilon,\eta,\sigma$ are arbitrary signs.

\fig{The four types of folds of the long edge separating two
triangles in the $d$-FCH model: complete fold, no fold, 
right angle up, right angle down.
The figure is drawn in the space spanned by
$\vec{f_i},\vec{f_j},\vec{f_k}$ of \twoposd.}{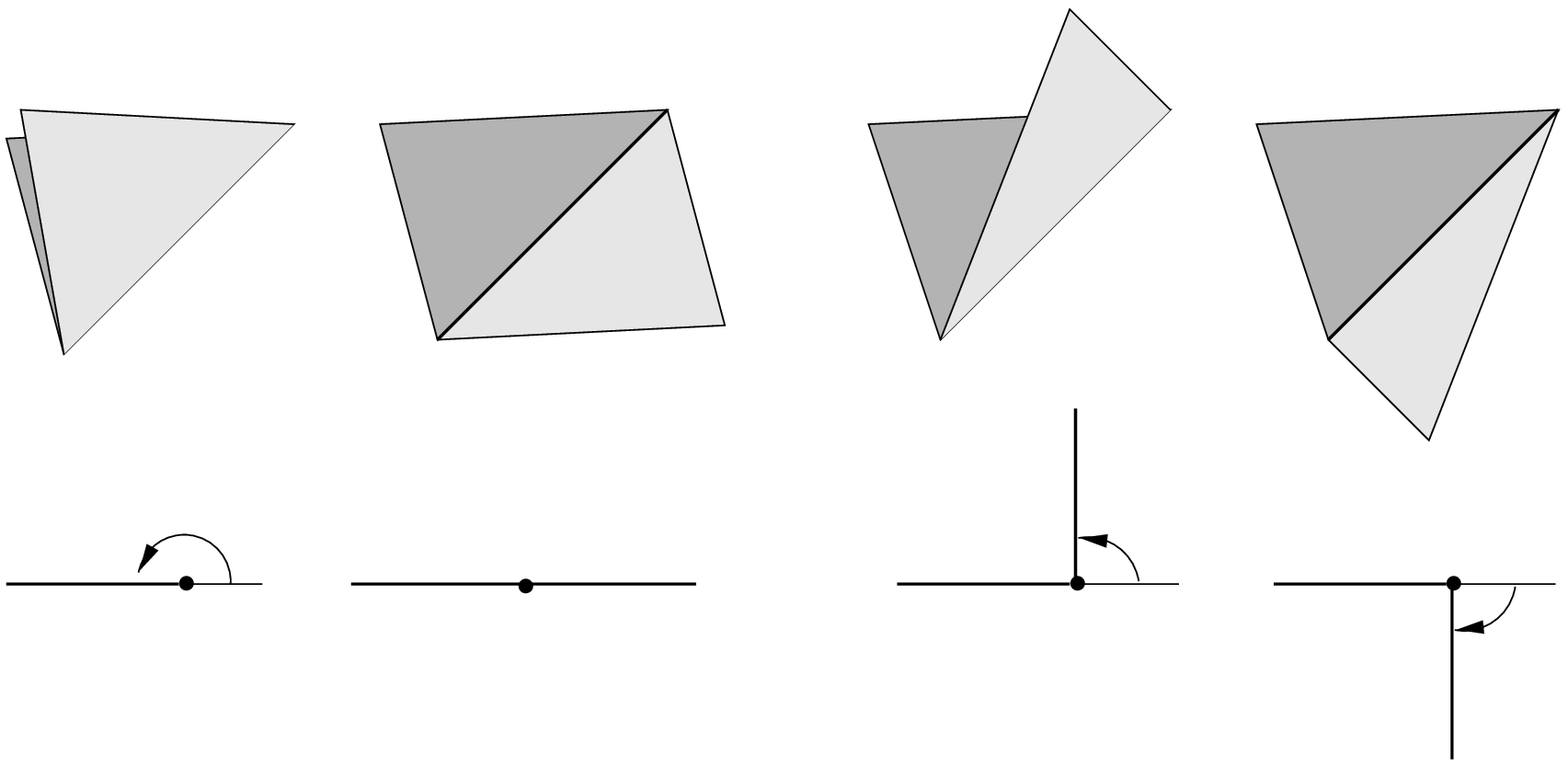}{8.cm}
\figlabel\foltree

This gives rise to essentially four types of foldings for the long edge,
as depicted in Fig.\foltree.
If $i=k$, the long edge
is completely folded ($180^\circ$) when $\sigma=\eta$ and
unfolded when $\sigma=-\eta$. If $i\neq k$, the long edge is
always folded at a right angle, either up when $\sigma=\eta$ ($+90^\circ$),
or down if $\sigma=-\eta$ ($-90^\circ$).

\subsec{Vertex Model}

Let us again consider the square lattice $S$ whose vertices are 
the 8-valent vertices
of the square-diagonal lattice. The edges of $S$ are the long edges 
of the original lattice.
According to the previous section, these edges are folded into edges
of the form 
\eqn\mafodfc{ \rho({\vec t})~=~ \epsilon \vec{f_i} }
Each such image is therefore characterized by a color 
$i\in\{1,2,...,d\}$ and a sign $\epsilon\in \{-1,1\}$.
Let us examine the possible relative images of the long edges in two
triangles sharing a short edge. There are basically only two possibilities
\eqn\bastopo{ \figbox{2.5cm}{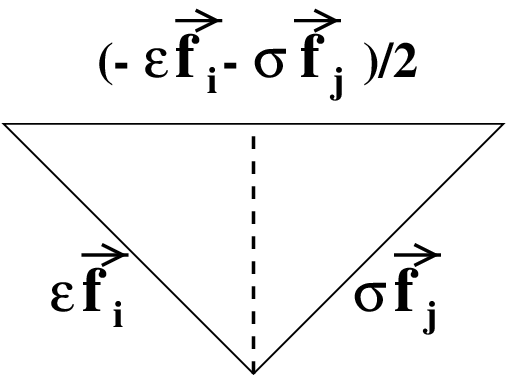} \quad {\rm or}
\quad \figbox{2.5cm}{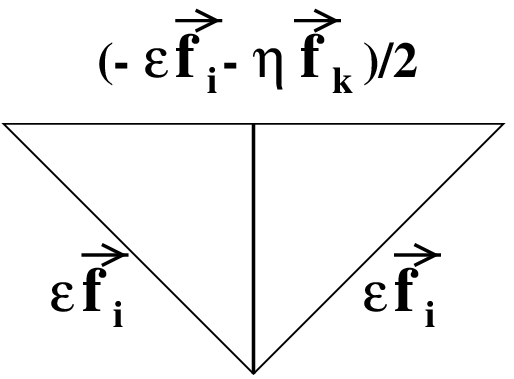} }
to accomodate the face rule \rulfac, where $i\neq j$, $i\neq k$, and
$\epsilon,\sigma,\eta$ are arbitrary signs.  
We have indicated only the value of the central short edge image, as
the remaining ones are immediately deduced from the face rule \rulfac.
As a corollary, we see that the short edges are always either completely
folded (solid line in \bastopo), or unfolded 
(dashed line in \bastopo). 
Together with Fig.\foltree, 
this shows a complete parallel between the $d$-HCD and $d$-FCH models: we see
that the roles of the long and short edges are exchanged in the two
models, at least as far as the types of folding are concerned.
Note however that there are twice as many short edges as long edges 
in the square-diagonal lattice, we therefore expect a difference in
the folding entropies of the two models.

\fig{The allowed configurations of long edges around
a face of $S$ in the $d$-FCH model. 
We have indicated the degeneracy of each face configuration
and the corresponding Boltzmann weight. The inner short lines
represent the state of the corresponding short edges,
namely completely folded (solid line) or 
unfolded (dashed line).}{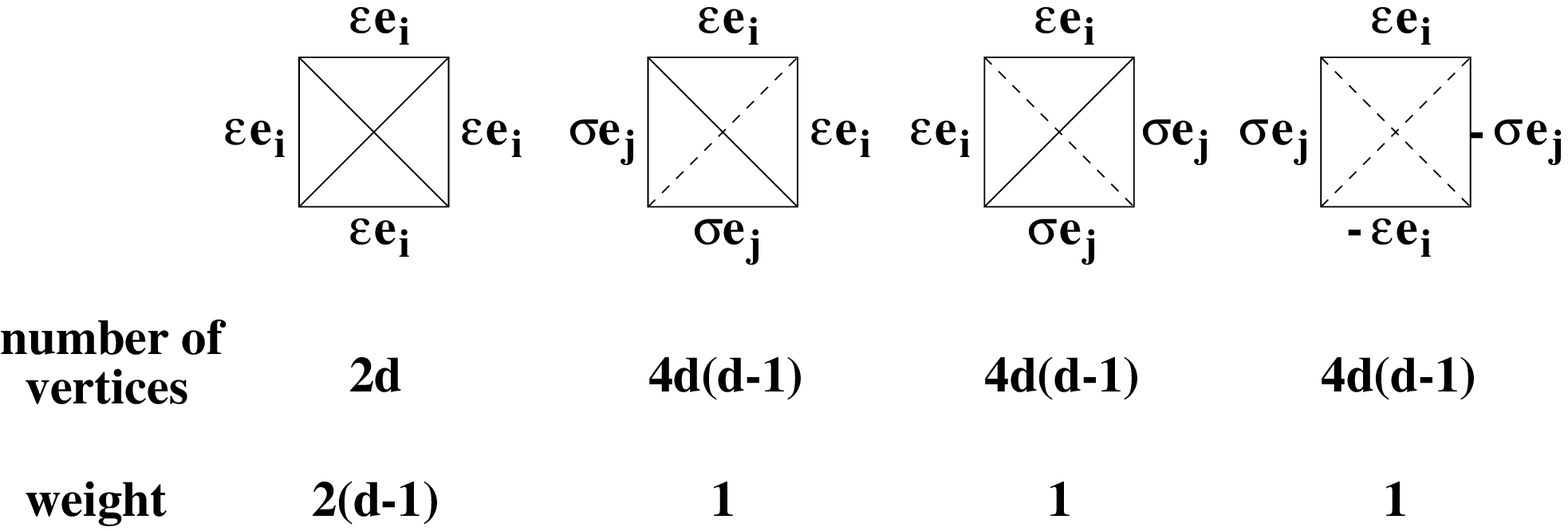}{9.cm}
\figlabel\verd

Let us now investigate the possible configurations of the images of the
long edges around a face of $S$.
Using \bastopo\ and proceeding by inspection, we have found the 
face configurations displayed in Fig.\verd, together with their
degeneracy and attached Boltzmann weight. From \bastopo, we find
a unique image for each internal short edge in the three last cases
of Fig.\verd\ (hence a weight 1), whereas in the first case much
freedom is left: indeed, the inner edge images may take
values $(-\vec{f_i}\pm \vec{f_k})/2$, for any value of $k\neq i$,
the same for all edges,
and with alternating signs as we move around the face.
This results in $2\times (d-1)$ possible inner values, hence the
corresponding Boltzmann weight of Fig.\verd.
Let us count these allowed configurations. From Fig.\verd,
we count
\eqn\counver{ W_d~=~ 2d+3\times 4d(d-1)~=~ 2d (6d-5) }
distinct vertices.

\fig{The allowed vertices of the $W_d$ Vertex model, with the corresponding
Boltzmann weight. The edge variable is defined modulo $2d$ and we have
$a\neq b$ mod $d$.}{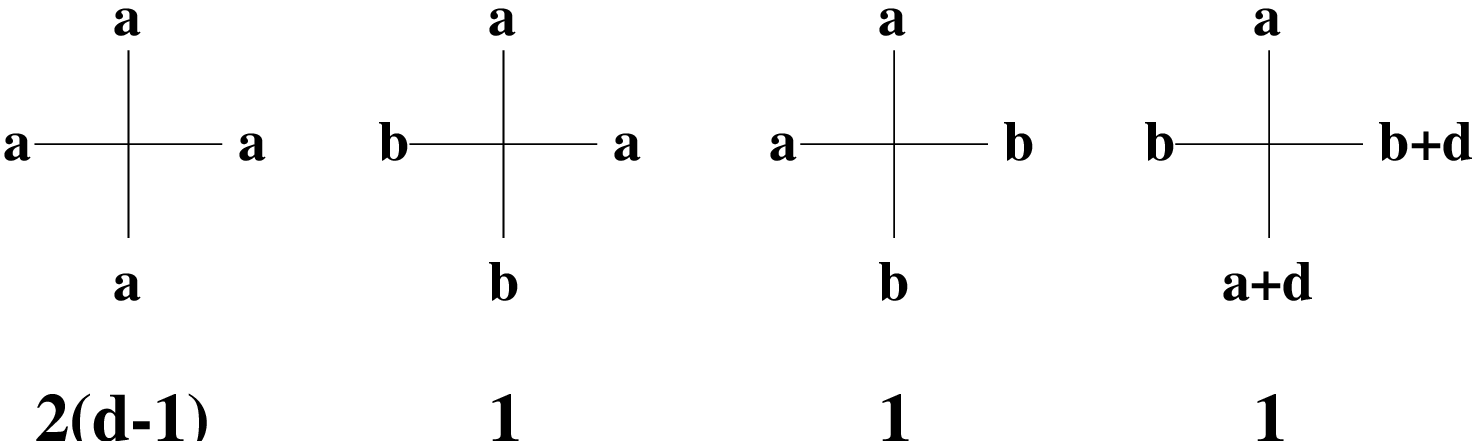}{7.cm}
\figlabel\verdet

The face configurations of Fig.\verd\ may be recast into the vertex
configurations of Fig.\verdet\ of the dual square lattice $S^*$, in which each
edge is assigned an image $a \in \IZ_{2d}$, with the correspondence
\eqn\corwd{(+,\vec{f_i})~ \to~ a=i-1\qquad {\rm and}
\qquad (-,\vec{f_i})~\to~ a=d+i-1}
for $i=1,2,...,d$. 

When $d=2$ we recover the $W_2=28$ Vertex model, with the correct
assignments of Boltzmann weights of Fig.\verset, up to the 
correspondence \corwd\  between the edge variables.
This shows how our two ($d$-HCD and $d$-FCH) $d$-dimensional foldings 
agree with the unique model for $d=2$ in quite different ways, 
providing for two different points of view on the $d=2$ model.
We will elaborate on this in a later section.

The present $W_d$ Vertex model appears to be much simpler than the
$V_d$ Vertex model for the $d$-HCD folding. Indeed, the
edge variables may take only $2d$ values (to be compared to
the $2d(d-1)$ values in the other model), 
and the number of distinct vertices
\counver\ is only a quadratic polynomial of $d$ 
(to be compared with the quartic polynomial \totver). 
The only price to pay for this simplification is the Boltzmann weight 
$2(d-1)$ for completely folded faces.

The transfer matrix for the $W_d$ Vertex model is again sparse, with a
ratio $[(6d-5)/(4d^2)]^N$ of non-vanishing elements.
It gives easier access to numerical calculations, at least for
small enough values of $d$.

\subsec{Loop Model}

\fig{The four types of colored vertices for the
$d$-FCH folding model. The solid and dashed lines 
stand for any two distinct colors in $\{1,2,...,d\}$.
The signs are conserved along colored lines,
except for the last vertex, where they are reversed.}{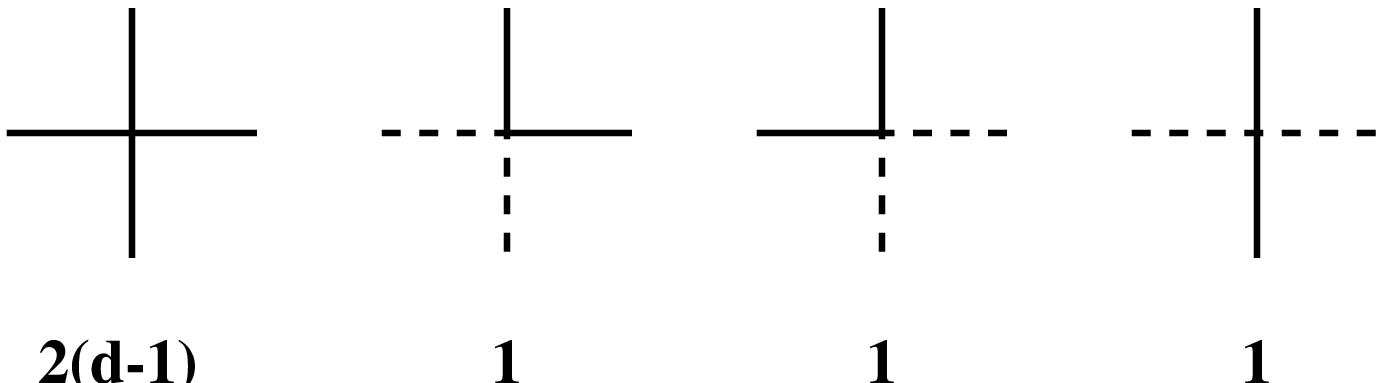}{8.5cm}
\figlabel\colvert

The $W_d$ Vertex model of the previous section can be refined as follows.
Let us disentangle the sign and color variables on each edge of $S^*$,
and represent only the color of the edge, by painting it accordingly.
This leaves us with only
the four types of vertices depicted in Fig.\colvert. 
the signs must now be conserved or reversed along colored lines
according to Fig.\verd. The sign is actually conserved in all cases,
except when the two lines of different color cross each other (last,
completely unfolded case of Fig.\verd), in which case it is reversed.
These colored lines form loop-like clusters, along which the 
signs are entirely
determined by their value on one of the edges of the cluster. Moreover,
all values are compatible, as there are always an even number of sign
reversals (due to an even number of crossings with lines of other colors)
along a loop.

\fig{A sample coloring configuration for the $d$-FCH folding 
model, for $d=4$.}{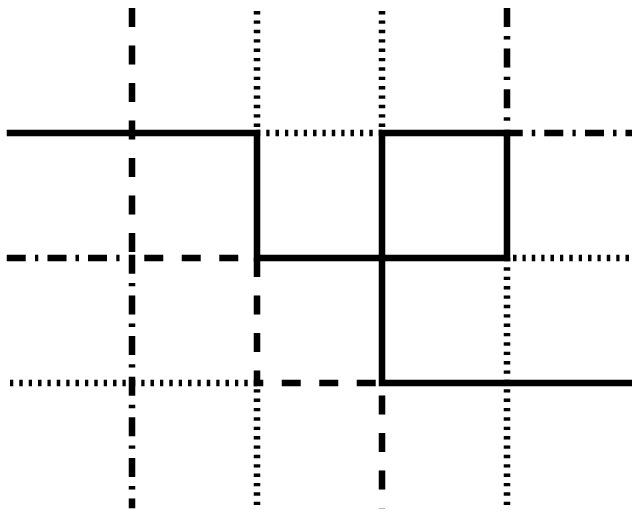}{7.cm}
\figlabel\sacod

Hence on top of the Boltzmann weights indicated in Fig.\colvert, we must
include a weight $2$ per loop-like cluster of given color.
A sample configuration of the model for $d=4$ 
is displayed in Fig.\sacod\ for illustration. 
With free boundary conditions, this configuration would receive the
weight $2^7 \times 2(4-1)=1536$, as $7$ colored clusters are formed,
and one vertex is of the first type of Fig.\colvert.

\subsec{Estimates for the d-FCH Folding Entropy}

The loop model of previous section permits to derive 
lower bounds 
on the partition function $Z_{FCH}^{(d)}$ of the
$d$-FCH folding model, henceforth on the $d$-FCH folding entropy. 

We first note that there are exactly $2d$ 
fundamental configurations of the loop model
with minimum energy (i.e. maximum Boltzmann weight), namely
those obtained with only the first vertex of Fig.\verdet, that is 
where all the edges are painted with the same color.
The contribution to the partition function of each of these 
groundstates is $[2(d-1)]^n$ for a portion of $S^*$ with $n$ vertices.
Using Fig.\verd, these correspond to completely folded configurations
of all the short edges.

\fig{A few local excitations of the all-black groundstate
of the $W_d$ Vertex model. We have represented a single
excitation (a), formed by a loop of any of the $(d-1)$ other
colors and a pair of neighboring excitations (b) with necessarily identical
colors. The single excitation (a) suppresses 4 black vertices
of the groundstate, whereas the pair suppresses 6 of them.}{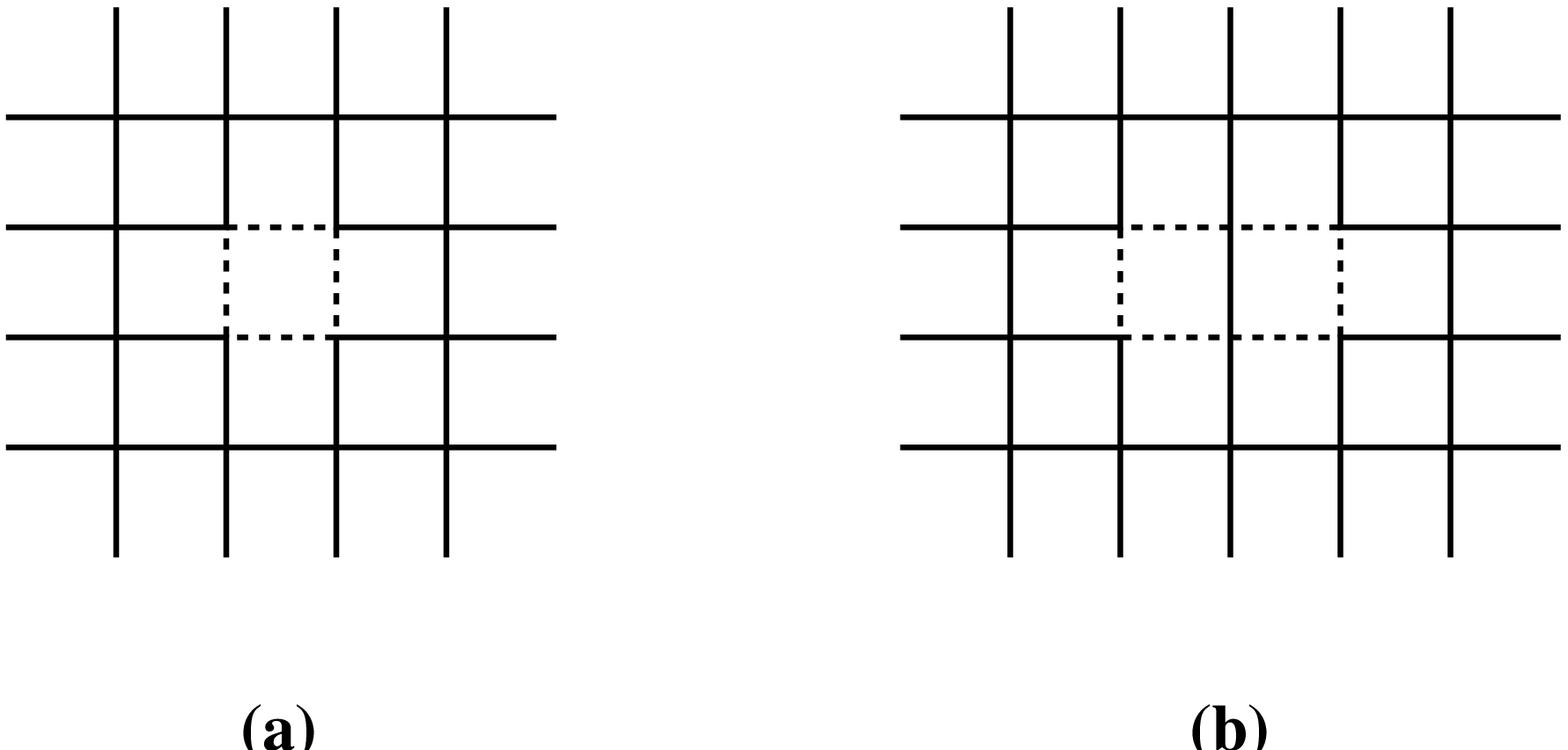}{8.cm}
\figlabel\coldex

Let us fix such a groundstate, say of color $1$ (black). 
Consider the local excitation
obtained by forming a minimal loop of another color, as depicted in
Fig.\coldex (a). The loop suppresses
4 vertices of the first type in Fig.\colvert, but introduces an 
extra weight $2(d-1)$ for the extra loop and the choice of $d-1$
other colors to paint it. Hence the local excitation contributes an overall
relative weight of $1/(2(d-1))^3$.  These excitations do interact whenever
they occupy neighboring faces of $S^*$. For instance,
as shown in Fig.\coldex (b), two such neighboring
excitations are forced to have the same color, and form only one extra loop,
whereas they suppress 6 vertices of the first type in Fig.\colvert.
Hence, instead of the expected relative weight $1/(2(d-1))^6$, they only
contribute for $(2(d-1))/(2(d-1))^6=1/(2(d-1))^5$. This propagates to
higher numbers of excitations. In general, the partition function 
is underestimated if we neglect these interactions. This 
provides us with an exact lower bound
\eqn\lobex{ Z_{FCH}^{(d)} ~\geq ~ \bigg[2(d-1)(1+{1 \over 
(2(d-1))^3})\bigg]^n }
hence
\eqn\slobex{ s_{FCH}^{(d)} ~\geq~ {1\over 4} {\rm Log}\bigg(2(d-1)
(1+{1 \over (2(d-1))^3})\bigg) }

\par\begingroup\parindent=0pt
\leftskip=1cm\rightskip=1cm\parindent =0pt
\baselineskip=11pt
$$\vbox{\font\bidon=cmr8 \def\bidondon{\bidon} \bidondon \offinterlineskip
\halign{\tv \quad # \tv
& \hfill \ #
& \hfill # \tv \cr
\noalign{\hrule}
\tvi  $d$ \hfill & entropy \hfill & p. func. \hfill\cr
\noalign{\hrule}
\tvi 2 &  .20273255 & 1.224744  \cr
\tvi 3 &  .35044964 & 1.419705  \cr
\tvi 4 &  .44909460 & 1.566892  \cr
\tvi 5 &  .52034819 & 1.682613  \cr
\tvi 6 &  .57589615 & 1.778723  \cr
\tvi 7 &  .62137130 & 1.861478  \cr
\tvi 8 &  .65985542 & 1.934512  \cr
\tvi 9 &  .69320821 & 2.000122  \cr
\tvi 10 & .72263580 & 2.059855  \cr
\noalign{\hrule}
}}$$
{\bf Table V:} Lower bound on the $d$-dimensional FCH folding
entropy, from \slobex. We display the dimension $d$, the lower bound
on the folding entropy and the corresponding lower bound
on the partition function per triangle.
\par
\endgroup\par

The values of the lower bound \slobex\ for the $d$-FCH model
are displayed in Table V. 
Note that \slobex\ consists of the two leading orders of
the Mayer expansion of $Z_{FCH}^{(d)}$ in terms of the local excitations
described above.
The expansion however is quite involved, as the excitations interact
between nearest neighbours (two faces sharing an edge)
and second-nearest neighbors (two faces sharing a vertex) as well.
In the latter case, we have to use extra caution and 
distinguish between the cases when the two excitations have
the same or different colors.
Up to order $2$ in these excitations, we find
\eqn\mayZ{ \big(Z_{FCH}^{(d)}\big)^{1\over n}~=~
u(1+{1 \over u^3}+{5 \over u^5} -{8 \over u^6}
+O({1\over u^7}))}
where we have set $u=2(d-1)$. Note the bad apparent convergence at $d=2$
($u=2$), where this yields $z\simeq 1.233$, way below our best
estimate $1.258$. 
It appears that for $d=2$ the Mayer expansion \mayer\ around
a very different groundstate is much more accurate. So, 
although both have the same contribution $2^n$ to the partition function,
the latter groundstate yields better corrections.
This competition between different groundstates could yield
an interesting phase diagram for a fully interacting model.

We expect however the expansion  \mayZ\ to behave better
for larger values of $d$, for which the existence of a groundstate
generalizing that leading to \mayer\ is not clear.

Comparing \slobex\ to the large $d$ estimate 
\shcd\ for the HCD model, we conclude that the two $d$-dimensional
folding models (HCD and FCH) agree in the large $d$ limit.

\subsec{Numerical Study}

In view of the previous sections, the $W_d$ Vertex model 
is a straightforward generalization of the $28$ Vertex model 
we have already studied.
A direct adaptation of our programs permits to calculate
the largest eigenvalue of the transfer matrix for a few values
of $d$. 

Like in the $d=2$ case, we use the method of iteration of 
the action of the transfer matrix $T$ on a given vector, which we normalize
at each step. To construct $T$, we simply list its $(6d-5)^n$ non-zero
elements, for fixed boundary conditions (say $=0$) on the leftmost edge of
the row. Each such element has a unique pair of (row,column) indices
in $T$: indeed, the type of vertex in Fig.\verdet\ is fixed uniquely
whenever the east, north and south edge variables are given. 
Like in the $d=2$ case, this linear description of the matrix gives
access to large sizes. We list our results in Table VI below, for
fixed boundary conditions ($=0$) on both ends. 

\par\begingroup\parindent=0pt
\leftskip=1cm\rightskip=1cm\parindent =0pt
\baselineskip=11pt
$$\vbox{\font\bidon=cmr8 \def\bidondon{\bidon} \bidondon \offinterlineskip
\halign{\tv \quad # \tv
& \hfill \ #
& \hfill #
& \hfill # &\tv 
\quad # \tv
& \hfill \ #
& \hfill #
& \hfill # \tv \cr
\noalign{\hrule}
\tvi $\ $ & $n$ & $\lambda_{\rm max}$ \hfill & 
$\nu_n$ \hfill &
$\ $ & $n$ & $\lambda_{\rm max}$ \hfill & 
$\nu_n$ \hfill \cr
\noalign{\hrule}
\tvi $\ $ & 1 & 4.0000000 &         &  & 1 & 6.0000000 & \cr
\tvi $d=3$ & 2 & 16.271109 & 1.420166 & 
$d=4$ & 2 & 36.170708 & 1.566936\cr
\tvi $\ $ & 3 & 67.253188 & 1.425850 &   
& 3 & 218.18854 & 1.567179\cr
\tvi $W_3=78$ & 4 & 280.96435 & 1.429666 & 
$W_4=152$ & 4 & 1316.2983 & 1.567222\cr
\tvi $\ $ & 5 & 1185.8001 & 1.433309 &   
& 5 & 7941.2213 & 1.567231\cr
\tvi $\ $ & 6 & 5063.2725 & 1.437490 &   
&   &           &        \cr
\noalign{\hrule}
\tvi $\ $ & 1 & 8.0000000 &         &  & 1 & 10.000000 & \cr
\tvi $d=5$ & 2 & 64.127762 & 1.682631 & 
$d=6$ & 2 & 100.10133 & 1.778729\cr
\tvi $\ $ & 3 & 514.66822 & 1.683140 &   
& 3 & 1002.5120 & 1.778944\cr 
\tvi $W_5=250$ & 4 &  4131.2200 & 1.683207 & 
$W_6=372$  & 4 & 10040.689 & 1.778969\cr
\tvi  $\ $ & 5 & 33162.598 & 1.683226 &   &   &           &         \cr
\noalign{\hrule}
}}$$
{\bf Table VI:} Numerical results for the  transfer matrix
of the $W_d$ Vertex model with fixed boundary conditions ($=0$ on both ends).
We list the dimension $d$, the
length $n$ of the row, the largest eigenvalue $\lambda_{\rm max}$,
and the  sequence 
$\lambda_{\rm max}^{1/(4n)}$, and 
$\nu_n=(\lambda_{n+1}/\lambda_n)^{1/4}$,
converging to the partition function per triangle.
\par
\endgroup\par

These values are extrapolated to the following
\eqn\folextra{ \matrix{
s_{FCH}^{(3)}&=.378... &  z_{FCH}^{(3)}&=1.47...\cr
s_{FCH}^{(4)}&=.4493... &  z_{FCH}^{(4)}&=1.5672...\cr
s_{FCH}^{(5)}&=.5207... &  z_{FCH}^{(5)}&=1.6832...\cr
s_{FCH}^{(6)}&=.57603... &  z_{FCH}^{(6)}&=1.77896... \cr }}
for the entropy $s_{FCH}$ and the partition function per triangle
$z_{FCH}$.
Note the excellent agreement with the lower bounds of Table V, for
$d=4,5,6$.
We expect the Mayer expansion \mayZ\ to be an excellent approximation to 
the partition function per site for all $d\geq 4$. 

\par\begingroup\parindent=0pt
\leftskip=1cm\rightskip=1cm\parindent =0pt
\baselineskip=11pt
$$\vbox{\font\bidon=cmr8 \def\bidondon{\bidon} \bidondon \offinterlineskip
\halign{\tv \quad # \tv
& \hfill \ #
& \hfill # \tv \cr
\noalign{\hrule}
\tvi  $n$ \hfill & $\lambda_{max}$ \hfill & $\nu_n$ \hfill\cr
\noalign{\hrule}
\tvi 1 & 5.000000 & \cr
\tvi 2 & 24.21917  & 1.48353  \cr
\tvi 3 & 119.4202  & 1.49014  \cr
\tvi 4 & 582.3144  & 1.48600  \cr
\tvi 5 & 2812.592  & 1.48247  \cr
\tvi 6 & 13498.13 &  1.48010 \cr
\noalign{\hrule}
}}$$
{\bf Table VII:} Numerical results for the  transfer matrix
of the $W_3$ Vertex model with mixed boundary conditions ($=0$ on
one end, free on the other) for the 3-dimensional FCH folding.
We have represented the size $n$ of the row, the largest eigenvalue
and the ratio $\nu_n=(\lambda_{n+1}/\lambda_n)^{1/4}$, which
converges to the partition function per triangle. 
\par
\endgroup\par

For $d=3$, we observe a much slower convergence (see Table VI).
To obtain better results, we have also calculated the 
largest eigenvalue of the transfer matrix $T$ in the case of mixed boundary
conditions, fixed at $0$ on one end, and free on the other.
The results are displayed in Table VII.
Upon extrapolation, this leads to a more precise result for the 
$3$-dimensional FCH folding entropy and partition function per triangle:
\eqn\finesdtr{ s_{FCH}^{(3)}~\simeq~.3854...  
\qquad z_{FCH}^{(3)}~\simeq~1.470.. }

\newsec{Other Compactly Foldable Lattices}

\subsec{Classification of Compactly Foldable Lattices}

We would like to briefly address the following question: can one
classify the inequivalent two-dimensional lattices which are
compactly foldable onto themselves in two dimensions?
By this we mean that all maps $\rho$ preserving the
face rule \rulfac\ and the lengths of edges, and with values in $\IR^2$
actually have their image included in the original lattice.
Another formulation is the existence of a map $\rho$ whose
image is a single face of the lattice
(the lattice is then completely foldable onto that face).

We already know of three examples: the square lattice (and its trivial
variation the rectangular lattice), the regular triangular lattice
and the square-diagonal lattice introduced in this paper.
\fig{The square-diagonal lattice as the superposition
of two square lattices $S$ and $S'$. The edges of $S$ are represented
in solid lines, those of $S'$ in dashed lines.}{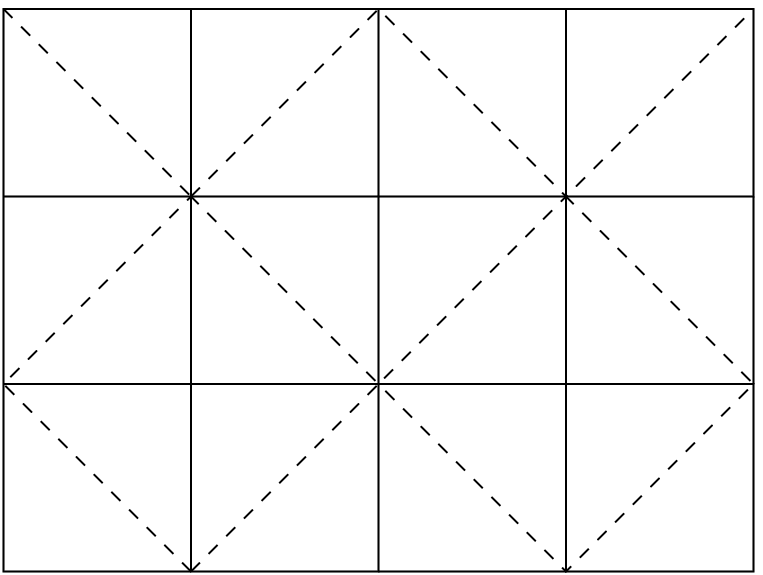}{5.cm}
\figlabel\tilt
One way to look at the square-diagonal lattice
is rather as the superposition of two square lattices (see Fig.\tilt),
say $S$
and $S'$, such that $S'$ is a dilated version of $S$ by a factor
$\sqrt{2}$ as well as rotated by $45^\circ$, so that the vertices of
$S'$ coincide with half of the vertices of $S$, forming a checkerboard.
\fig{The dilation/rotation transformation on the triangular lattice
gives rise to another triangular lattice represented in dashed lines.
The superposition of the two forms a new foldable
lattice, with long, medium  and short edges.}{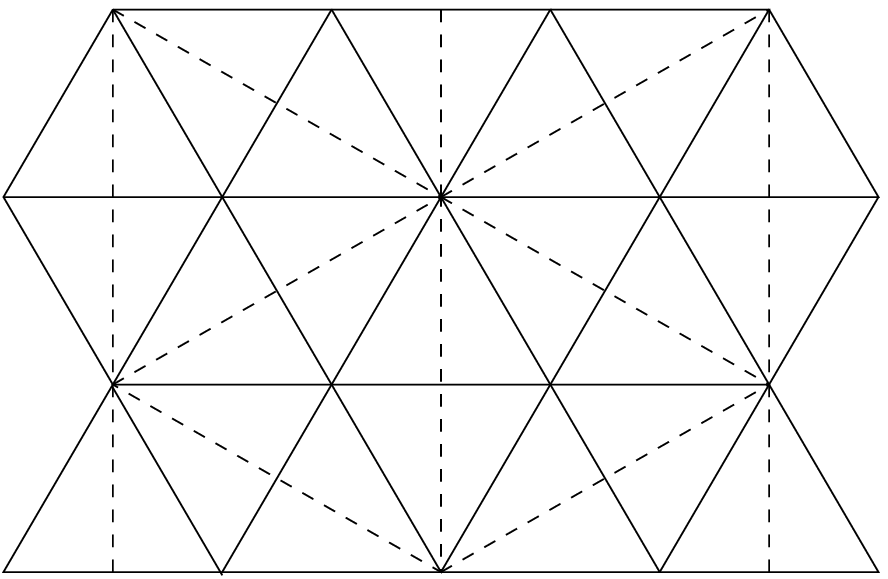}{5.cm}
\figlabel\tiltri
This suggests to apply the same type of transformation
(dilation/rotation) to the triangular lattice.
The only non-trivial possibility is a dilation by a factor of
$\sqrt{3}$, and a rotation by $30^\circ$, displayed in Fig.\tiltri.
This gives rise to a new foldable lattice in two dimensions,
by superposition of the two triangular lattices, called the 
double-triangular lattice.
We expect its entropy of two-dimensional folding to be larger than
that of the square-diagonal model. This model will be studied in
the next section.

This can be shown to actually exhaust all the possibilities for
compactly foldable two-dimensional lattices.
As a side remark, the generating function for compactly foldable
random triangulations of arbitrary genus was obtained in \DEG.

\subsec{Two-Dimensional Folding of the Double-Triangular Lattice}

The double-triangular lattice of Fig.\tiltri\ has three types
of edges: long, medium, short of respective lengths $2$, $\sqrt{3}$, $1$. 
Each triangular face has one edge of each type.
As usual, we introduce tangent vectors along these edges, with compatible
orientations throughout the lattice, so that the face rule \facerul\
is satisfied around each triangular face.

\fig{The diamond lattice formed by the long edges
of the double-triangular lattice (solid lines), and its dual,  
the Kagom\'e lattice (dashed lines).}{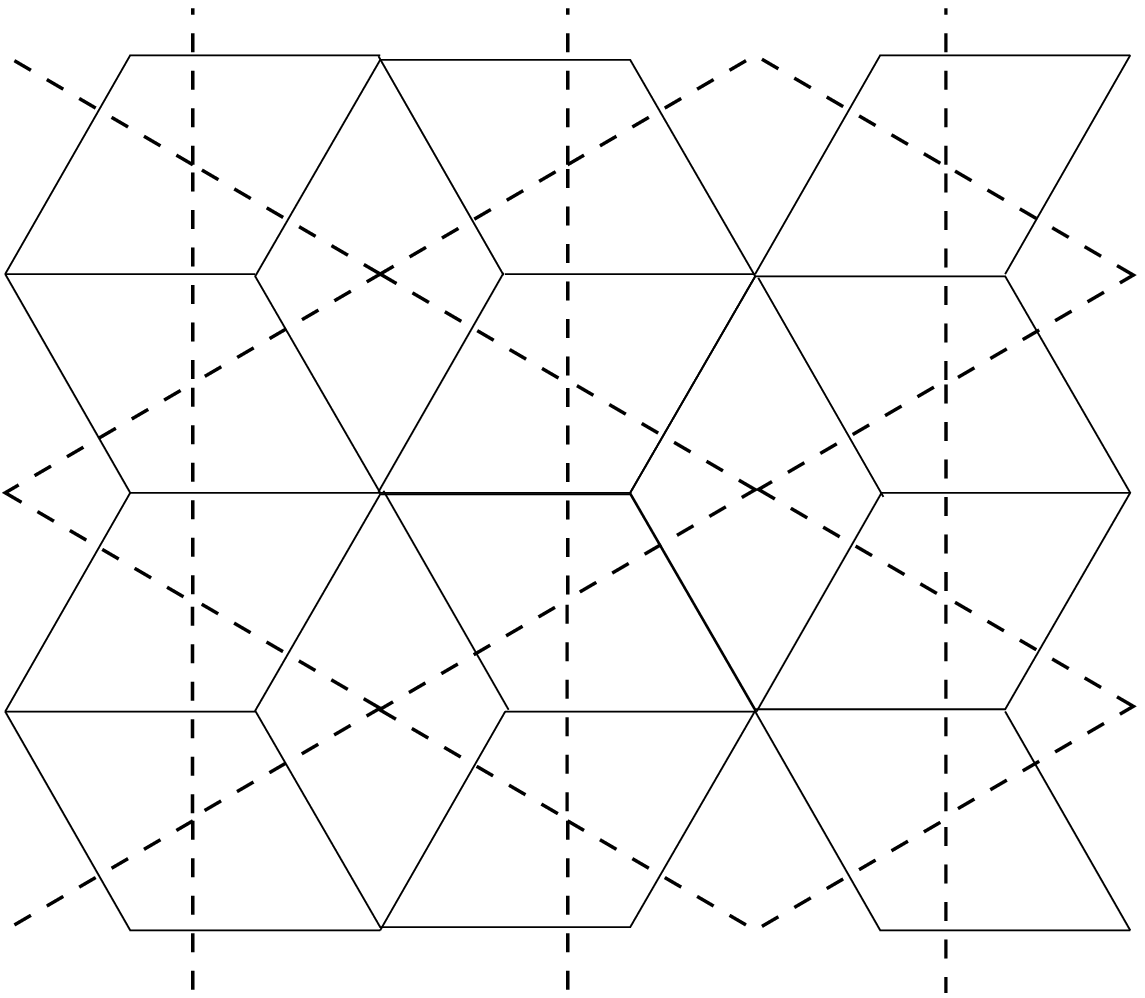}{8.cm}
\figlabel\diam

A folding configuration of the lattice is a continuous map
$\rho$ of these tangent vectors to the plane, such that the face
rule \rulfac\ is satisfied around each elementary triangular face. 
Let us now first concentrate on the long edges of the 
double-triangular lattice.
They form the diamond-lattice, represented in solid
lines in  Fig.\diam, which is dual to
the Kagom\'e lattice, represented in dashed lines in the
same figure. 

By inspection, it is easy to see that
the images of the long edge vectors may only take one of the six
values $\pm \vec{e_1}$, $\pm \vec{e_2}$, $\pm \vec{e_3}$, where
the $\vec{e_i}$ are three fixed vectors of length $2$ with vanishing
sum (hence forming angles of $120^\circ$).
As before, writing these images as
\eqn\imagtt{ \rho(\vec{t})~=~ \epsilon \vec{e_i}} 
this suggests to attach a color $i=1,2,3$ to each long edge, and
a sign $\epsilon=\pm 1$.

\fig{The four possible configurations of long edges around
a diamond-shaped face. We have represented in dashed lines
the (medium or short) unfolded inner edges, and in solid lines the folded
inner edges.
We have also indicated the attached Boltzmann weights. 
The color indices take the values $i,j=1,2,3$,
with $i\neq j$, and $\epsilon,\sigma$ are arbitrary 
signs.}{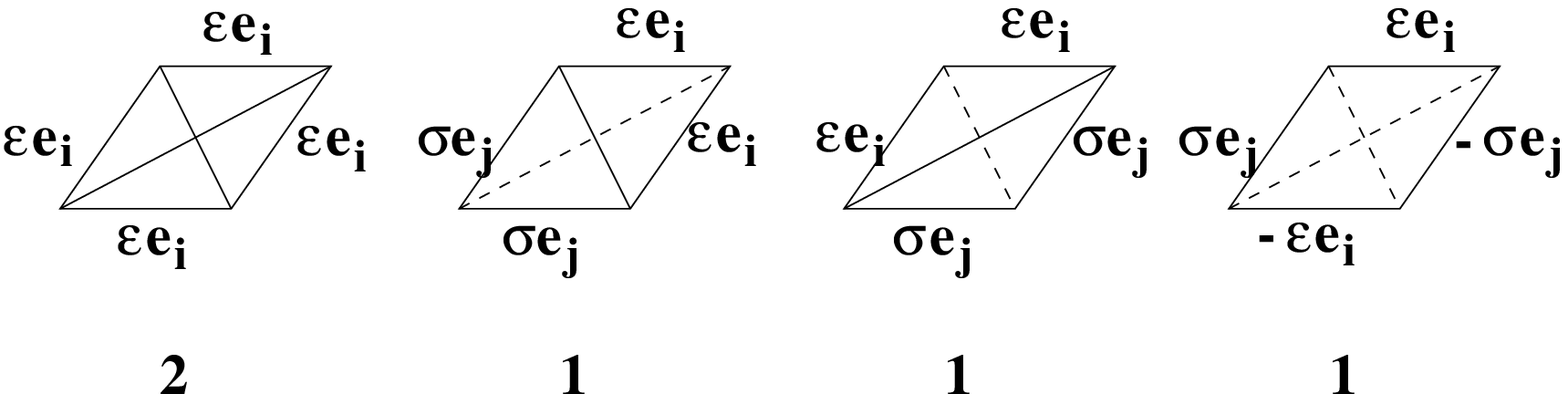}{10.5cm}
\figlabel\fourdiam

In a way very similar to the square-diagonal case, the long edges
around any diamond-shaped face of Fig.\diam\ may only take the
four possible relative values depicted in Fig.\fourdiam,
according to the folding state of the inner short and medium edges. 
Note the Boltzmann weights, $1$ for the last three cases of Fig.\fourdiam,
as the inner edges are entirely fixed, and $2$ for the first case, as
we have two choices for the
inner short edges $\vec{s}=\epsilon \vec{e_k}/2$,
$k\neq i$, which then fix all other inner edges. Each long edge may take
$6$ values. This gives a total of $78$ distinct
possible diamond face environments. 
These turn out to be in one-to-one
correspondence with the $78$ cases of Fig.\verd, when $d=3$, 
but with different Boltzmann weights.
This gives a remarkable relation between a two-dimensional
folding problem and a three-dimensional one.  

\fig{The allowed vertices of the
tri-colored cluster model on the Kagom\'e lattice, equivalent to
the two-dimensional DT folding problem. The solid and dashed
lines stand for any two distinct colors among $\{1,2,3\}$.
We have indicated the Boltzmann weight under the three
corresponding vertices.
Each colored cluster has a fugacity $2$.}{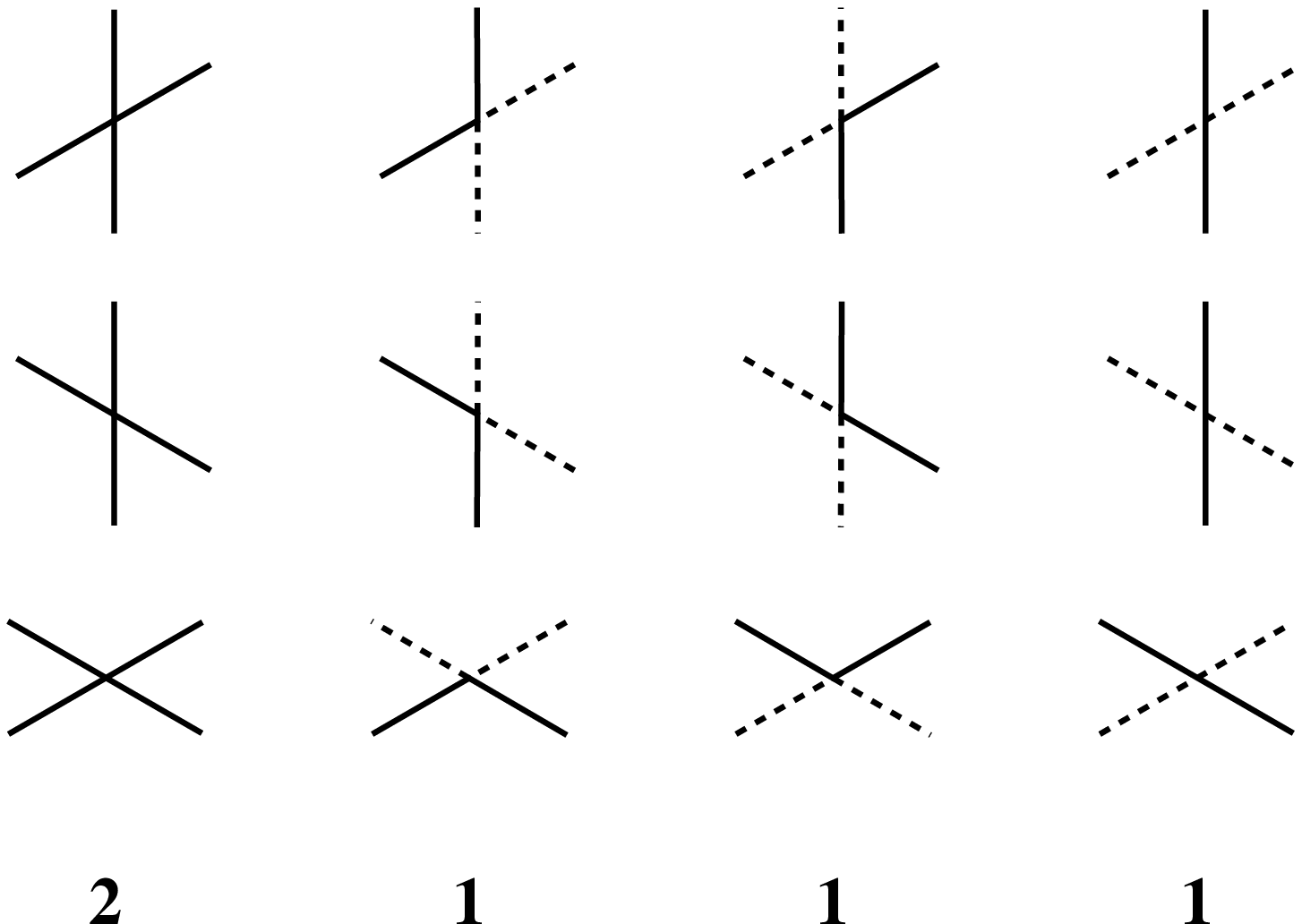}{8.cm}
\figlabel\kag

Like in the d-FCH folding case, we may now rephrase the 
folding problem as a $78$ Vertex model on the edges of
the dual Kagom\'e lattice of Fig.\diam, in which each edge may take
a value $a \in \IZ_6$, and with the vertices derived from
Fig.\fourdiam.
The transformation into a colored cluster model performed in Sect.7.3
is still valid here. Hence the model is equivalent to a colored 
cluster model with the vertices of Fig.\kag, 
similar to that of Fig.\colvert, with $3$ edge colors,
and with a weight $2$ per vertex of the first type, and an
extra Boltzmann weight of $2$ per cluster.
Note that the three different types of vertices of the Kagom\'e lattice
are not distinguished in the present model.

This permits to derive a simple lower bound on the two-dimensional
folding entropy $s_{DT}$ of the double-triangular lattice. 
Indeed, let us start from one of the $3$ groundstates of the vertex
model, in which all the edges are painted with the
same color. There are $N/4$ such vertices, corresponding to $N/4$ diamond
faces in the dual, each of which contains 4 triangles, for a total 
number of $N$ triangles. We therefore obtain the lower bound 
\eqn\stargrou{ Z_{DT}~\geq~ 2^{N/4} }
for the partition function of the two-dimensional double-triangular
lattice model.

This can be easily improved by considering the local excitations
of Sect.7.4. In the present case, these correspond to creating a
minimal triangular loop of one of the two other colors than that of
the groundstate. Such a loop suppresses $3$ vertices of weight $2$ each,
and contributes for an extra weight $2$ for the two 
choices of color, and the weight $2$ for the loop created.  
Each such local excitation contributes therefore a relative weight
$4/2^3=1/2$ to the partition function. As before, neglecting the
contact interactions between these excitations, we get an exact lower
bound for the partition function 
\eqn\lobdt{ Z_{DT}~\geq ~ 2^{N/4} (1+{1\over 2})^{N/6} }
where we have identified the total number $N/6$ 
of excitable triangles.
This gives the lower bound
\eqn\lobsdt{ s_{DT}~\geq~{1\over 4}{\rm Log}\, 2 + 
{1\over 6} {\rm Log}\big(1+{1\over 2}\big)~=~ .2408...}
or a partition function per triangle of at least $1.2723...$

Two excitations sharing a vertex interact as follows. If the
two excited triangles have the same color, they receive a weight
$2$ for the choice of color, $2$ for the cluster formed, and 
$5$ black vertices are erased, but one of the other color is created,
hence a total contribution
$2\times 2/2^4=1/4$. 
If their colors are distinct, they receive a weight $2$ for the choices
of this pair of colors, $2\times 2$ for the two created loops,
and $5$ black vertices are suppressed, hence a total weight $2^3/2^5=1/4$.
Two neighboring excitations therefore contribute a total of $1/4+1/4=1/2$
instead of the expected $1/4$. These may take any of the $N/4$ positions
available for their common vertex on the Kagom\'e lattice. 
Up to two excitations, the Mayer
expansion of the folding entropy therefore reads
\eqn\maytrok{ s_{DT}~\simeq~{1\over 4}{\rm Log}\, 2 +
{1\over 6} {\rm Log}\big(1+{1\over 2}+{3\over 8}\big)~=~.278...}
or a partition function per triangle $z_{DT}\simeq 1.320...$. 
 
\subsec{d-dimensional Folding of the Double-Triangular Lattice}

We now briefly address the question of the
d-dimensional foldings of the double-triangular lattice.

\fig{The unit cells of the two possible $3$-dimensional lattices
onto which that of Fig.\tiltri\ can be folded. Both are
decorations of the $3$-dimensional FCC lattice, whose unit
cell is an octahedron. We have represented the vertices added
to the original FCC unit cell by filled dots.}{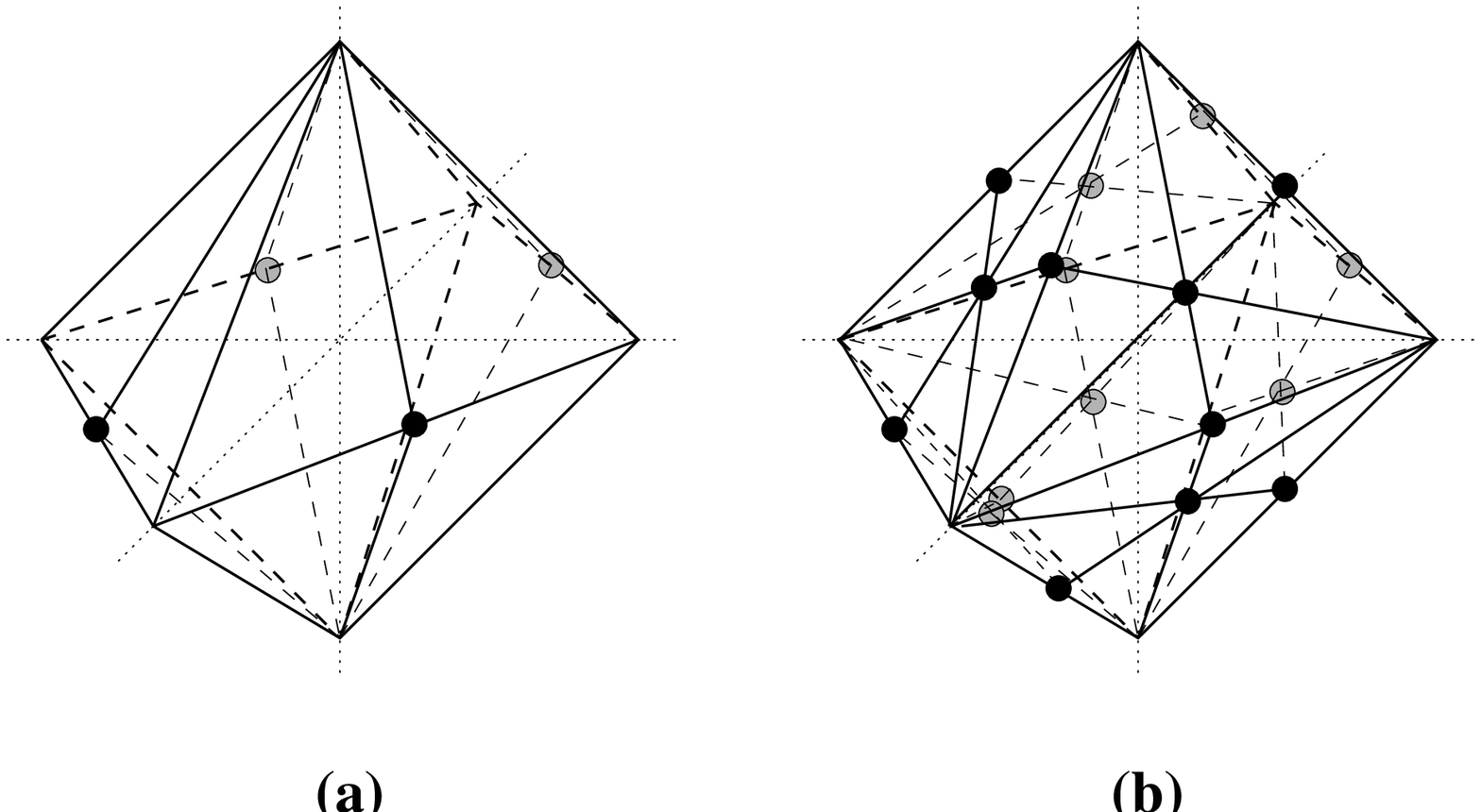}{9.cm}
\figlabel\dtri

The reader will convince himself that there are again two possibilities
here, according to whether we privilege long or short edges.
The first possibility consists in starting from the $d$-dimensional FCC
lattice of \DGGF, whose unit cell is made of a polytope, with
$2d$ vertices at the points $(0,0,..,0,\pm 1,0,..,0)$ in $\IR^d$,
with edges joining any pair of these points with non-zero coordinate
in different places,
and regular triangular faces joining any triple of such points.
Next we
introduce new vertices in the middle of one third of all
edges, which we join to the opposite vertices of all faces adjacent
to that edge, thus constructing exactly one height on each triangle.
We do this in a coherent way, by successively sending the height from
a triangle to its neighbors by rotating it around any of the two
opposite edges of its face. In the end, each section of the lattice
by a plane containing one of its faces must be the
lattice of Fig.\tiltri. The octahedral unit cell of the
$3$-dimensional version is represented in Fig.\dtri (a).

The other possibility also starts from the $d$-dimensional FCC lattice.
We next simply add a central vertex in the middle of each face and
of each edge, and join them by drawing the three heights of each
triangular face of the FCC lattice. The unit cell of this lattice
is represented in Fig.\dtri (b) for $d=3$.

Note that the heights in these two constructions play exactly
the role of the
diagonals in the case of the square-diagonal lattice folding.
Both models lead naturally to higher vertex models on the Kagom\'e lattice.

\newsec{Discussion and Conclusion}

\subsec{Folding Entropies}

In this paper, we have found various estimates for the entropy of 
folding of the square-diagonal lattice into target spaces of
various dimensions.
Let us compare the results we have obtained to another 
folding problem, that of the triangular lattice in two and 
three dimensions \DG\ \DGGF. 

In two dimensions, the partition function per triangle for the
folding of the triangular lattice was
obtained in \DG, with the exact result
\eqn\stri{ z_{T}^{(2)}~=~ {\sqrt{3} \over 2\pi} 
\Gamma(1/3)^{3/2} ~=~1.20872...}
The square-diagonal folding partition function per triangle
\eqn\zptrid{ z_{SD}~=~1.2586...}
is substantially larger. In addition, our model presents
two different types of edges, to which we can 
attach different bending energies. Doing so should produce
an interesting phase diagram and hopefully confirm the
first order folding transition of \DGT.
The double-triangular folding partition function per triangle
\eqn\dtrent{ z_{DT}~\simeq~1.32...}
is even larger, as expected.

In three dimensions, the best numerical estimate of \DGGF\ for the
partition function per triangle of the triangular lattice folding reads
\eqn\tras{z_{T}^{(3)}~=~1.43(1) }
Our present estimates 
\eqn\prestim{\eqalign{
z_{HCD}^{(3)}~&=~1.35(1)\cr 
z_{FCH}^{(3)}~&=~1.470(1)\cr} } 
lie on both sides of \tras. The main advantage of the FCH folding
model is that it deals with much smaller transfer matrices,
allowing in principle for better estimates.
Again, the introduction of bending rigidity (in the form of a
bending energy, e.g.) should lead to an interesting phase diagram.
Eq.\prestim\ is evidence of the difference between the
two (HCD and FCH) $3$ dimensional foldings of the square-diagonal lattice.
However, we also noted that the two models should agree better 
and better as $d$ increases.

A last remark is in order, regarding the relation between the 
$d$-dimensional HCD and FCH models.
Let us consider the following restriction of the
$W_d$ Vertex model for the $d$-FCH model (called rFCH model), 
in which the first 
vertex of Fig.\verdet\ is assigned a weight $0$. This forbids
the formation of colored clusters, by suppressing the first vertex
of Fig.\colvert.  We are therefore left with a model of
colored lines, with colors $i=1,2,...,d$, which either avoid (by being
reflected)
or cross each other on the vertices of $S^*$, according to whether 
we have the second, third or fourth vertex of Fig.\colvert.
This is therefore a model of dense colored loops, each affected with a
weight $2$ (for the two choices of signs along each loop).
Its partition function reads
\eqn\restup{ Z_{rFCH}^{(d)}~=~\sum_{{\rm dense}\ {\rm colored}\atop
{\rm loops}} 2^{N_1+N_2+...+N_d} }
This is strongly reminding of the partition function for the
$d$-HCD model \pfdcol. Note however two discrepancies. Firstly the 
coloring rules are different. 
Although the gluing rules along the edges of the squares in the $d$-HCD
model \proref\ are the same as the vertex rules in the rFCH model 
(crossing or reflection), they do not take place on the same sites 
(vertex v/s mid-point of edge).
This is probably not a very important difference. The main one is
that there are $N/4$ vertices in the rFCH model (for $N$ triangles),
and $N/2$ edges in the $d$-HCD model. So even if the partition functions
were identical, we would get 
$s_{rFCH}^{(d)}=s_{HCD}^{(d)}/2$ in the thermodynamic
limit. From \prestim, we see then that when $d=3$
only a small portion of the
FCH entropy is taken care of by the restricted model. 
Note finally that
the restriction imposes that {\it at most} one set of two parallel
short edges can be folded on each face of $S$: this rigidifies the
faces of $S$ only partially, and the latter can be thought of as 
a random distribution of exactly one  diagonal on each
face of a square lattice $S$, each of which is 
either folded or not, hence this looks like a 
randomized version of the square-diagonal lattice, in which the
former long edges play the role of short edges, and pairs of
parallel former short edges play the role of long edges.

\subsec{Bending Rigidity}

A natural extension of our models would be the introduction of
a bending rigidity, namely of an energy making folding more or
less difficult. 

In the two-dimensional Square-Diagonal
folding problem, this can be implemented by
attaching a Boltzmann weight $e^{K_i}$, per unfolded 
long ($i=1$) or short ($i=2$) edge, while completely
folded edges receive weights $e^{-K_i}$ respectively.
Positive values of $K_i$ will favor a completely flat groundstate,
whereas negative values of $K_i$ will favor a completely folded
groundstate. 

Interestingly, if we choose $K_1<<0$ and $K_2=0$, the system will 
tend to maximize its number of folded long edges. This corresponds 
to either of the two groundstates of Sect.2.3
which have led us to \minofo, with a free energy 
\eqn\enerone{ -{f_{-0}\over kT}~=~-{K_1  \over 2}+{1\over 4} {\rm Log}2 }
per triangle (there are $N/2$ long edges for $N$ triangles).

If instead we choose $K_1=0$ and $K_2<<0$, the system will
have a tendency to have all its short edges folded, hence, in
terms of the $W_2=28$ Vertex model, the first vertex of
Fig.\verset\ will be favored, namely all edges will have the same
value of $a$. 
The corresponding four groundstates are those of Sect.7.4
which have led us to \lobex, for $d=2$.   
The corresponding free energy per triangle reads
\eqn\enerto{ -{f_{0-}\over kT}~=~-K_2+{1\over 4}{\rm Log}2 }
(there are $N$ short edges for $N$ triangles).

More generally, the situation $K_2=0$, $K_1$ arbitrary can still be
described within the framework of the $28$ Vertex model of Fig.\verset.
Indeed, the first vertex receives a Boltzmann weight $2e^{-4K_2}$
(all four inner short edges are folded), the last one
a weight $e^{4K_2}$ (all four inner short edges are flat)
and the two remaining vertices receive a weight $1$ 
(two short edges are folded, two are flat).
This model will be studied elsewhere.

The most general model with $K_1$ and $K_2$ arbitrary forces us to 
go back to the initial degrees of freedom of the folding model.
As we have just seen, the 28 Vertex model allows only for including
a bending rigidity for short edges. Introducing a rigidity
for long edges as well will introduce an interaction between neighboring
vertices, and force us to disentangle the first  vertex in Fig.\verset,
and to rather use the $V_2=32$ Vertex model formulation of Sect.6.4
(see \totver). This new interacting model has a more sophisticated
transfer matrix, as the vertices interact with their nearest neighbors. 

The two particular limits $K_1\to \infty$, $K_2$ finite,
or $K_2\to \infty$, $K_1$ finite 
of the general model can be easily solved.
They correspond to the complete flattening of one type of edges, thus
forming rigid square faces, which can be folded along the other type 
of edges. This is then equivalent to the folding of the square lattice,
whose very simple phase diagram was studied in \DGT.
The main result was a first order phase transition between a completely
flat phase and a completely folded one. What makes the model very simple
is the absence of folding entropy, as all the folds initiated on the
boundary of the studied domain propagate all the way through the lattice
along straight lines.
Note that there are $N/2$ flattened squares in the first case (each
square is made of 2 triangles), whereas there are $N/4$ such squares
in the second case (each square is made of four triangles). 
The corresponding free energies read
\eqn\enersq{\eqalign{
K_1\to \infty\ : \ 
-{f_{K_1,K_2}\over kT}~&\simeq~{K_1\over 2}+|K_2| \cr
K_2\to \infty\ : \
-{f_{K_1,K_2}\over kT}~&\simeq~K_2+{|K_1|\over 2} \cr}}
  
\subsec{Fluid Membrane Folding}

We would like to finally comment on a fluid version of the
models studied in this paper. A fluid membrane can be modelled 
by a random tessellation of a Riemann surface. This amounts to
replacing the regular lattices representing the tethered membrane
by random tessellations with the same type of tiles
as the regular lattice, but with an arbitrary connectivity at each vertex.

In the case of the square-diagonal lattice, let us choose for tiles
the square faces of $S$, each made of $4$ triangles. In the $W_d$ Vertex
formulation of the $d$-FCH folding, each of these faces is replaced by a 
4-valent vertex, with edge values $a\in\IZ_{2d}$. This suggests to
introduce a fluid version thereof, in which the lattice is replaced 
by $\phi^4$ (4-valent) graphs, with edge variables $a\in \IZ_{2d}$.
A simple way of representing the partition function of such a model
is to consider a Hermitian multi-matrix integral 
\eqn\multima{ Z(N,g)~=~\int dM_0...dM_{2d-1} 
e^{-N\,{\rm Tr}(V(M_0,...,M_{2d-1}))} }
where $M_0$, $M_2$, ...,$M_{2d-1}$ are $N\times N$ Hermitian
matrices, integrated wrt the standard Haar measure, and
\eqn\poten{\eqalign{
V(&M_0,...,M_{2d-1})~=~\sum_{i=0}^{2d-1}({1\over 2g}M_i^2
+{d-1\over 2}M_i^4)\cr
&+{1\over 4}\sum_{0\leq i\neq j \leq d-1} (M_i^2M_j^2+2M_{i+d}^2M_j^2
+M_{i+d}^2M_{j+d}^2+4M_iM_jM_{i+d}M_{j+d}) \cr}}
where the quartic terms reproduce the vertices of Fig.\verdet.
Moreover, the measure of integration is normalized so that 
$\lim_{g\to 0}Z(N,g)=1$.
The perturbative expansion of ${\rm Log}Z$ as a power 
series of $g$ and $N$
reads ${\rm Log}Z=\sum_{h,n\geq 0} N^{2-2h} g^n Z_{n,h}$, where $Z_{n,h}$
is the partition function for the $W_d$ Vertex model on the set of
connected tessellations of surfaces of genus $h$ with $n$ squares. 
Unfortunately, even for $d=2$ the solution of the model \multima\ is not
known.
\medskip
\noindent{\bf Acknowledgements}

We thank K. Hallowell for interesting discussions. This work was partially
supported by NSF grant PHY-9722060.

\listrefs
\bye